%% file: osorio_final.tex
\documentclass[]{aa}

\usepackage{graphicx}

\begin{document}

\title{A CCD Imaging Search for Wide Metal-Poor Binaries\thanks{Based on observations made with the IAC80 telescope operated on the island of Tenerife by the Instituto de Astrof\'\i sica de Canarias in the Spanish Observatorio del Teide; also based on observations made with the 2.2\,m telescope of the German-Spanish Calar Alto Observatory (Almer\'\i a, Spain), the William Herschel Telescope (WHT) operated on the island of La Palma by the Isaac Newton Group in the Spanish Observatorio del Roque de los Muchachos (ORM) of the Instituto de Astrof\'\i sica de Canarias; and the Telescopio Nazionale Galileo (TNG) at the ORM.}}

\author{M.\,R$.$ Zapatero Osorio
        \inst{1}
 \and
        E.\,L$.$ Mart\'\i n
        \inst{2}
 }

\offprints{M.\,R$.$ Zapatero Osorio, \email{mosorio@laeff.esa.es}}

\institute{LAEFF-INTA, P.\,O$.$ Box 50727, E--28080 Madrid, Spain          
   \and
       Instituto de Astrof\'\i{}sica de Canarias, E--38200 La Laguna, 
       Tenerife, Spain
 }

\date{Received; accepted }

\abstract{We explored the regions within a radius of 25\arcsec~around 473 nearby, low-metallicity G- to M-type stars using $(VR)I$ optical filters and small-aperture telescopes. About 10\%~of the sample was searched up to angular separations of 90\arcsec. We applied photometric and astrometric techniques to detect true physical companions to the targets. The great majority of the sample stars was drawn from the Carney--Latham surveys; their metallicities range from roughly solar to [Fe/H]\,=\,$-3.5$\,dex. Our $I$-band photometric survey detected objects that are between 0 and 5 mag fainter (completeness) than the target stars; the maximum dynamical range of our exploration is 9\,mag. We also investigated the literature and inspected images from the Digitized Sky Surveys to complete our search. By combining photometric and proper motion measurements, we retrieved 29 previously known companions, and identified 13 new proper motion companions. Near-infrared 2MASS photometry is provided for the great majority of them. Low-resolution optical spectroscopy (386--1000\,nm) was obtained for eight of the new companion stars. These spectroscopic data confirm them as cool, late-type, metal-depleted dwarfs, with spectral classes from esdK7 to sdM3. After comparison with low-metallicity evolutionary models, we estimate the masses of the proper motion companion stars to be in the range 0.5--0.1\,$M_{\odot}$. They are orbiting their primary stars at projected separations between $\sim$32 and $\sim$57000\,AU. These orbital sizes are very similar to those of solar-metallicity stars of the same spectral types. Our results indicate that about 15\%~of the metal-poor stars have stellar companions at large orbits, which is in agreement with the binary fraction observed among main sequence G- to M-type stars and T\,Tauri stars.  }

\titlerunning{Imaging Search for Wide Metal-Poor Binaries}
\authorrunning{Zapatero Osorio \& Mart\'\i n}

\maketitle

\section{Introduction}

Halo subdwarfs are metal-deficient ([Fe/H]\,$\le$\,$-1$\,dex), high-velocity ($v_{\rm tan}$\,$\ge$\,200\,km\,s$^{-1}$) stars. They belong to the oldest known galactic population and are subluminous with respect to main-sequence stars of the same mass. Traditionally, they have been identified using proper motion surveys (e.g., Giclas \cite{giclas71}) and, more recently, objective prism plates (e.g., Beers et al$.$ \cite{beers85}). Such surveys are magnitude-limited and consequently biased towards intrinsically brighter stars. Only recently a subdwarf sequence of very-low mass halo subdwarfs has been identified with a cutoff at M($V$)\,=\,14.5 and $(V-I)$\,=\,2.8\,mag (Monet et al$.$ \cite{monet92}). The bolometric luminosities, effective temperatures and metallicities of these stars, and the faint cutoff, are an important constraint to stellar structural and evolutionary models. Gizis \& Reid (\cite{gizis97}) have stressed the importance of wide binary systems for checking the metallicity scale of low-mass dwarfs.

In 1991 we started a {\sc ccd}-based imaging survey aimed at finding wide low-mass companions to halo subdwarfs. Wide binary systems are like small clusters because they offer us the possibility of studying two stars of different mass, but with the same age, distance and chemical composition. Early results of our survey were reported in Mart\'\i n \& Rebolo (\cite{martin92}) and Mart\'\i n et al$.$ (\cite{martin95}).  In this paper, we present the full survey, including the discovery of companions to 13 metal-deficient dwarfs. Spectroscopic observations of the low-mass secondaries have been obtained for eight of them. We derive spectral types for these new objects ranging from esdK7 to sdM3.

\begin{figure}
\centering
\includegraphics[width=8.5cm]{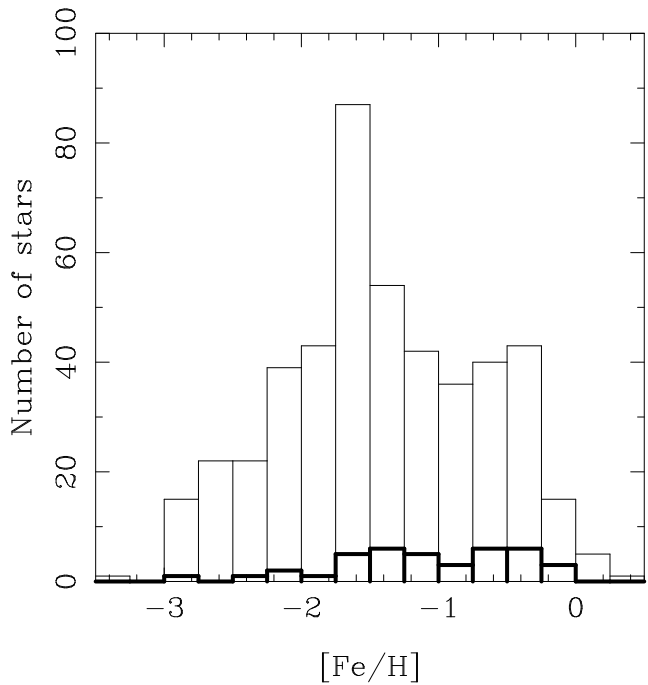}
\caption{The metallicity distribution of our entire sample is plotted as a thin line. The metallicity distribution of the stars with wide companions is displayed with a thick line. The bin width represents a change of 0.25\,dex in the metal logarithmic abundance.}
\label{metal} 
\end{figure}

\section{Sample selection}

We selected 473 G-, K- and M-type stars with known low-metallicities for our astrometric/photometric search. They were drawn from the surveys by Hartwick et al$.$ (\cite{hartwick84}), Laird et al$.$ (\cite{laird88}), Schuster \& Nissen (\cite{schuster89}) and Carney et al$.$ (\cite{carney94}). Chemical composition, as inferred from the [Fe/H] measurements given by the previous references, Ryan \& Norris (\cite{ryan91}) and Cayrel de Strobel et al$.$ (\cite{cayrel92}), is between solar abundance (0.0\,dex) and $-$3.5\,dex. The method of Carney et al$.$ (\cite{carney94}), which is the source of the great majority of our targets, to derive metal abundances is the best fit between observed spectra and a grid of synthetic spectra at the same effective temperatures but with varying metallicites. We will refer to these metallicity estimates as [Fe/H] throughout this paper. The distribution of metallicities is shown in Fig.~\ref{metal}: about 50\%~of the sample has a metal abundance between one and two orders of magnitude more depleted than the Sun, $\sim$30\%~of the stars have metallicities between 0.0 and $-$1.0\,dex, and the remaining $\sim$20\%~of the stars are metal depleted by more than a factor of 100. 

\begin{figure}
\centering
\includegraphics[width=8.5cm]{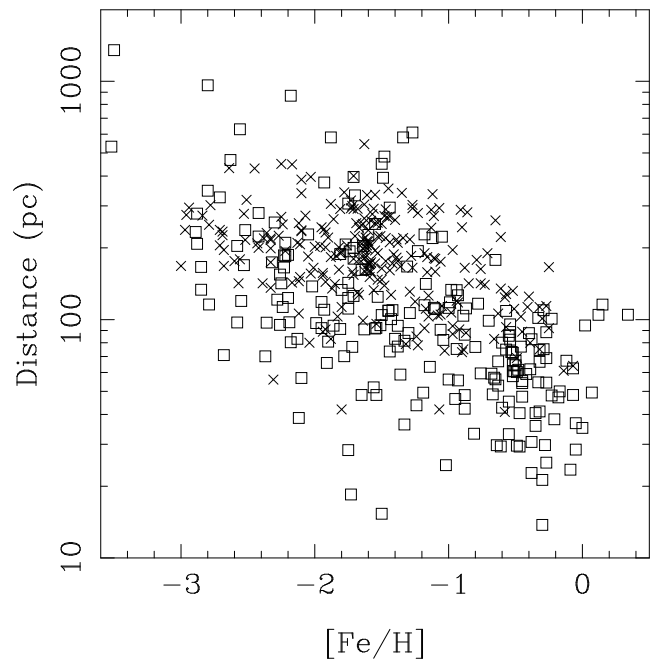}
\caption{Distance and metallicity of our target stars. Hipparcos stars (46\%~of the sample) are plotted as open squares, while crosses stand for stars with distances from Carney et al$.$ (\cite{carney94}).}
\label{dist} 
\end{figure}

Additionally, the sample comprises high proper motion stars (typically $\mu$\,$\ge$\,0.1\arcsec/yr), and with few exceptions, optical magnitudes are brighter than 14.5\,mag. The distance to the targets is between 15 and 1000\,pc in all cases, except for one Hipparcos star, as shown in Fig$.$~\ref{dist}. About 46\%~of the sample stars have Hipparcos parallaxes (Perryman et al$.$ \cite{perryman97}). Interestingly, Fig$.$~\ref{dist} also depicts the relation between metallicity and distance for nearby stars: in general terms, the most metal-depleted stars are located farther away than the most metal-rich stars. The lower envelope to this relation is given by: log\,$d$\,(pc)\,=\,1.146\,$-$\,0.268\,[Fe/H]. We only used Hipparcos data to derive this equation.

The list of all of our targets appears in Table~\ref{photlog} (the complete Table is available in electronic form). Stars are ordered by their increasing Giclas designation (Giclas \cite{giclas71}). Those stars with no Giclas name are listed at the end of the Table. We also provide $V$ magnitudes in the second column. We remark that we did not apply any particular criteria for the object selection, except for the metal-poor nature of the stars. 

\begin{table*}
\caption{List of photometric and astrometric observations.\label{photlog}}
\begin{center}
\begin{tabular}{lrccrccrrrc}
\hline
\hline
Object      &$V_p$~ & Fil. & Tel.  & Exp. & FWHM    & Date & $\rho$~ & PA~ & $\Delta$mag & Notes \\
            &       &      &       & (s)  &(\arcsec)&      &(\arcsec)&(deg)&             &       \\
\hline		
G\,001-004    & 10.69 & $I$ & IAC80      &   35  &  1.9     & 01/08/94  & $\ge$25 &         &        &  \\
G\,001-012    & 13.31 & $I$ & IAC80      &  100  &  2.6     & 02/08/94  & 23.29   &   276.8 & 4.49   & 1\\
G\,001-013    & 15.23 & $I$ & IAC80      &  330  &  2.5     & 02/08/94  & 20.48   &   341.2 & 3.89   & 1\\
\hline
\smallskip
\end{tabular}
\end{center}
{\sc notes.--} This is an example of the Table. The complete Table is available in electronic form.
\end{table*}

\section{Observations and analysis}

\setcounter{table}{1}
\begin{table*}
\caption{Previously known proper motion companions.\label{oldbin}}
\scriptsize
\begin{center}
\begin{tabular}{llcrrrrrrrrrrr}
\hline
\hline
Companion             & Primary          & [Fe/H]&$V_{\rm p}$&$R_{\rm p}$&$I_{\rm p}$&$\Delta(V)$&$\Delta(R)$&$\Delta(I)$&$\rho$~~~& PA~~ & Sep. &$d$ (Hip)&$d^a$~\\
                      &                  &       &           &           &           &           &           &           &($''$)~  & (deg)& (AU) & (pc)    & (pc) \\
\hline
G\,009-003\,B$^b$     &  G\,009-003\,A    &         & 13.07 &       &       & 5.92 & 5.25 & 4.15 &  15.88 &  63.0 &      &       &     \\ 
G\,026-008\,B         &  G\,026-008\,A    & $-$0.32 & 10.47 &       &       & 4.70 & 4.16 & 3.26 &   8.48 & 306.0 &  636 &       &  75 \\
G\,026-010$^c$        &  G\,026-009       & $-$0.99 &  9.89 &       &       & 5.04 & 5.53 & 6.19 & 133.04 &  29.3 & 6572 &  49.4 &     \\ 
G\,027-008\,B         &  G\,027-008\,A    & $-$1.53 & 11.39 & 11.06 & 10.70 & 5.71 & 5.13 & 4.62 &   5.63 & 173.3 &  828 &       & 147 \\
LP\,581-80            &  G\,028-040       & $-$0.87 & 11.33 &       &  8.17 & 3.20 &      & 2.25 &  15.81 & 349.8 & 1873 & 111.5 & 114 \\
G\,032-046\,B$^d$     &  G\,032-046\,A    & $-$0.57 &  9.79 &       &  9.02 & 8.81 &      & 8.74 &  62.52 & 168.1 & 4679 &  74.9 &  59 \\ 
G\,061-023$^e$        &  G\,061-024       & $-$0.53 &  8.99 &       &       & 0.77 &      &      &  81.90 & 221.7 & 5987 &  73.1 &  69 \\ 
GJ\,516\,B            &  G\,063-036$^f$   &         & 11.38 &  9.84 &  8.84 & 0.3: &      & 0.3: &   2.7: &  40:  &   37 &  13.7 &     \\
BD\,+06\,2932\,B      &  G\,066-022       & $-$1.30 & 10.47 & 10.09 &  9.67 & 3.19 &      & 2.24 &   3.60 & 250.5 &  316 &  87.9 &  61 \\
G\,069-004\,N$^g$     &  G\,069-004       & $-$0.63 &  7.98 &       &  7.22 & 0.84 & 0.87 & 0.81 &   6.16 &  22.9 &  325 &  52.8 &  31 \\
G\,095-057\,B$^h$     &  G\,095-057\,A    & $-$1.02 &  8.15 &  8.22 &  7.32 & 0.61 & 0.53 & 0.44 &   7.25 &  54.3 &  178 &  24.4 &  22 \\
G\,128-064\,B$^i$     &  G\,128-064\,A    & $-$0.27 &  9.63 &       &  8.80 & 2.10 &      & 2.13 & 261.60 & 320.3 &23177 &  88.6 &  65 \\
G\,017-027$^j$        &  G\,153-067       & $-$1.65 &  9.63 &  9.14 &  8.70 & 4.25 &      &      &1170.70 &  36.7 &56544 &  48.3 &  35 \\
G\,171-050\,B         &  G\,171-050\,A    & $-$2.94 & 13.08 &       &       & 5.25 & 4.72 & 4.32 &  22.48 & 268.4 & 6632 &       & 295 \\
G\,173-003            &  G\,173-002       & $-$0.07 & 10.05 &       &       & 4.16 & 3.63 & 2.82 &  22.51 &  34.6 & 1440 &       &  64 \\
G\,194-037\,B         &  G\,194-037\,A    & $-$2.03 & 15.31 &       &       & 2.63 & 2.53 & 1.86 &  19.71 &  48.2 & 7825 &       & 397 \\
G\,197-050$^k$        &  G\,197-049       & $-$1.50 &  9.86 &  8.86 &  8.08 & 3.72 & 3.08 & 2.71 &  14.68 &   9.4 &  225 &  15.3 &     \\
G\,217-004\,B         &  G\,217-004\,A    & $-$0.40 & 11.46 &       & 10.64 & 6.85 & 6.06 & 4.98 &  12.36 & 200.8 & 1085 &  87.8 & 137 \\ 
G\,230-047\,B$^l$     &  G\,230-047\,A    & $-$0.39 & 10.12 &       &       & 5.01 & 4.49 & 3.63 &  11.63 & 267.3 &  930 &       &  80 \\
G\,230-049\,B$^m$     &  G\,230-049\,A    & $-$0.66 & 8.51  &       &  7.89 & 0.11 &      &      &   1.33 &  29.0 &   76 &  57.1 &  43 \\ 
G\,242-074\,B$^n$     &  G\,242-074\,A    & $-$0.78 &  9.78 &       &  9.09 & 5.1: &      & 4.08 &  19.50 & 336.1 & 2276 & 116.6 &  91 \\
G\,251-054\,B$^o$     &  G\,251-054\,A    & $-$2.36 & 10.07 &       &  9.37 & 4.60 & 4.25 & 4.05 & 110.59 & 208.5 &28288 & 255.8 &  59 \\
G\,252-049\,B$^p$     &  G\,252-049\,A    &$\ge-0.3$&  9.48 &       &  7.93 & 0.11 &      &      &   1.53 & 112.0 &   32 &  21.2 &     \\ 
G\,255-038\,B$^q$     &  G\,255-038\,A    &         &  9.78 &  9.12 &  8.57 & 3.78 &      & 2.36 &  14.34 & 324.2 &  644 &  44.9 &     \\
G\,262-022$^r$        &  G\,262-021       & $-$1.07 & 13.75 & 13.40 & 12.96 & 0.68 & 0.42 & 0.19 &  29.54 & 148.4 & 8685 &       & 294 \\
G\,266-029\,B$^s$     &  G\,266-029\,A    & $-$1.18 &  8.96 &       &  8.61 & 0.40 &      &      &   0.17 &       &   39 & 227.8 &     \\
LHS\,541              &  G\,273-001$^t$   & $-$1.40 &  8.16 &  7.87 &  7.54 & 8.16 & 7.60 & 6.67 &  18.57 &  52.0 & 1542 &  83.0 &  37 \\
BD+16\,2116\,B$^u$    &  BD+16\,2116\,A   & $-$0.05 &  7.48 &       &  6.77 & 1.8: &      &      &   1.2: & 256:  &   34 &  28.5 &     \\
BD+23\,2207\,B$^v$    &  BD+23\,2207\,A   & $-$0.38 &  5.80 &       &  5.23 & 5.8: &      & 4.33 &   7.72 & 302.7 &  175 &  22.7 &     \\ 
\hline
\end{tabular}
\end{center}
$^a$~Distance from Carney et al$.$ (\cite{carney94}), except for G\,176-046\,D.\\
$^b$~$V$ magnitude from Salim et al$.$ (\cite{salim03}).\\
$^c$~White dwarf. Metallicity from Cayrel de Strobel et al$.$ (\cite{cayrel97}).\\
$^d$~CCDM\,J00453+1421\,B.  \\
$^e$~$\Delta V$ and astrometry from Salim et al$.$ (\cite{salim03}). $I$ (G\,061-023)\,=\,9.03\,mag (Hipparcos).\\
$^f$~G\,063-036 is a low-metallicity, M-type star (Stauffer \& Hartmann \cite{stauffer86}).\\
$^g$~Hipparcos separation and position angle are 6\farcs193 and 23\,deg, respectively.\\
$^h$~Hipparcos separation and position angle are 7\farcs307 and 54\,deg, respectively.\\
$^i$~G\,128-064\,B is possibly HIP\,115671. $\Delta V$ and astrometry from Salim et al$.$ (\cite{salim03}). $\Delta I$ from Hipparcos.\\
$^j$~$\Delta V$ from {\sc simbad}. Astrometry from from Salim et al$.$ (\cite{salim03}). G\,153-067 is also known as G\,017-025.\\
$^k$~$\Delta R$ from Weis (\cite{weis91}). Metallicity from Alonso et al$.$ (\cite{alonso96}).\\
$^l$~CCDM\,J20415+5730\,B.\\
$^m$~Not resolved in our images. Separation, position angle and $\Delta V$ from the Hipparcos database.\\
$^n$~CCDM\,J00582+8007\,B.\\
$^o$~CCDM\,J08111+7954\,B. \\
$^p$~Not resolved in our images. Astrometry and $\Delta V$ from Hipparcos; metallicity from Reid et al$.$ (\cite{reid01}).\\
$^q$~CCDM\,J13349+7430\,B. Hipparcos separation and position angle are 14\farcs28 and 324\,deg, respectively. See notes to Table~\ref{photlog}.\\
$^r$~$\Delta V$ from {\sc simbad}.\\
$^s$~Not resolved in our data. Astrometry and $\Delta V$ from Allen et al$.$ (\cite{allen00}).\\
$^t$~G\,273-001 is an Hipparcos binary (see text and notes to Table~\ref{photlog}).\\
$^u$~Hipparcos separation of 1\farcs81 and PA\,=\,246\,deg. Metallicity from Zakhozhaj \& Shaparenko (\cite{zakhozhaj96}).\\
$^v$~Metallicity from Cayrel de Strobel et al$.$ (\cite{cayrel92}).\\
\end{table*}

\begin{table*}
\caption{New proper motion companions.\label{newbin}}
\begin{center}
\begin{tabular}{lccrrrrrrrccc}
\hline
\hline
Companion &[Fe/H]&$V_{\rm p}$&$R_{\rm p}$&$I_{\rm p}$&$\Delta(V)$&$\Delta(R)$&$\Delta(I)$&$\rho$~~~& PA~~ & Sep  &$d$ (Hip)& $d^a$~\\
          &      &           &           &           &           &           &           &($''$)~~& (deg)& (AU) & (pc)    & (pc)  \\
\hline
G\,009-047\,B$^b$ & $-$1.93 &  7.70 &       &  6.88 & 8.70 & 8.23 & 8.05 &  81.93 & 229.5 &30800 & 375.94 &  13 \\ 
G\,059-032\,B$^c$ & $-$0.23 &  9.04 &       &  8.10 &      &      &      & 112.03 & 108.7 & 5381 &  48.05 &  58 \\ 
G\,090-036\,B     & $-$1.62 & 12.69 &       &       & 3.44 & 2.97 & 2.61 &   1.67 &  47.4 &  369 &        & 221 \\
G\,093-027\,B$^d$ & $-$1.23 & 11.64 &       &       & 2.7: & 2.2: & 1.8: &   3.52 & 101.3 &  275 &        &  78 \\
G\,116-009\,B     & $-$1.46 & 14.34 &       &       & 4.00 & 3.57 & 3.03 &  10.08 &  90.5 & 2712 &        & 269 \\
G\,128-077\,B     & $-$1.36 & 13.36 &       &       & 2.87 & 2.05 & 1.89 &   9.18 & 138.3 & 1533 &        & 167 \\
G\,172-016\,B     & $-$1.64 & 10.97 &       &       & 5.85 & 5.34 & 4.84 &   8.30 & 154.5 &  822 &        &  99 \\
G\,176-046\,D$^b$ & $-$1.67 & 12.59 & 12.16 & 11.70 & 4.89 & 4.76 & 3.89 &   4.80 & 152.9 &  610 &        & 127 \\
G\,188-022\,B     & $-$1.45 & 10.05 &       &  9.55 & 6.93 & 6.21 & 5.45 &   5.08 & 308.3 &  553 & 108.81 &  89 \\
G\,204-049\,B     & $-$0.97 & 10.85 &       &       & 4.3: & 3.7: & 3.1: &   2.7: &  72:  &  235:&        &  87 \\
G\,214-001\,B     & $-$2.03 & 12.08 &       &       & 5.74 & 4.95 & 4.59 &   5.14 & 353.0 &  802 &        & 156 \\
G\,216-045\,B     & $-$0.52 & 11.24 &       &       & 8.56 & 7.49 & 6.22 &  31.45 &  43.9 & 4151 &        & 132 \\ 
G\,273-152\,B     & $-$0.60 & 10.10 &  9.65 &  9.23&$\le$5.7&4.42 & 3.57 &   3.84 & 277.6 &  307 &  80.06 &  65 \\
\hline
\end{tabular}
\end{center}
$^a$~Distance from Carney et al$.$ (\cite{carney94}), except for G\,009-047\,B.\\
$^b$~Distance and metallicity from Laird et al$.$ (\cite{laird88}).\\
$^c$~Astrometry from the Digital Sky Survey images.\\
$^d$~According to Carney et al$.$ (\cite{carney94}), it is a visual binary, but nothing is found in either {\sc simbad} or the archives. \\
\end{table*}

\subsection{Optical photometric and astrometric survey}

The great majority of the direct imaging observations was conducted with the 0.8\,m IAC80 telescope at the Observatorio del Teide during 1994, 1995, 1996, 2002, and 2003. The telescope was equipped with a Thomson 1024\,$\times$\,1024 {\sc ccd} camera, which provided a pixel projection onto the sky of 0\farcs4325. Further optical observations were carried out with the 1024\,$\times$\,1024 {\sc ccd} cameras of the 1\,m Jacobus Kapteyn Telescope (JKT) and of the 2.5\,m Nordic Optical Telescope (NOT) at the Observatorio del Roque de los Muchachos (ORM) during 1991, 1993, and 1994. The corresponding plate scales were 0\farcs30 (JKT) and 0\farcs14 (NOT). The most recent images taken in 2003 were obtained with the IAC80 telescope (same instrumental configuration as in the past) and the {\sc cafos} instrument attached to the 2.2\,m telescope at the Calar Alto Observatory (CAHA), providing a pixel size of 0\farcs53. Some observing nights were hampered by cirrus. Hence, we did not perform any absolute photometric calibration. The survey was entirely conducted in the $I$ filter because faint dwarf companions are significantly redder than their primary stars. Some targets were also imaged in the $V$ and/or $R$ filters to determine color differences between targets and companion candidates. This program was usually carried out as a backup observing program whenever the seeing was rather poor ($\ge$2\arcsec) or there were many clouds. 

Raw frames were bias-subtracted and flat-field corrected using packages running inside the {\sc iraf\footnote{IRAF is distributed by National Optical Astronomy Observatory, which is operated by the Association of Universities for Research in Astronomy, Inc., under contract with the National Science Foundation.}} environment. Flat-fields were usually collected during dawn and dusk. Individual exposures ranged between a few seconds and 600\,s, depending on the brightness of the targets and the telescope diameter. These integration times did not yield deep images. However, these data are intended to detect objects typically between 0 and 5\,mag fainter than the target stars. This is the dynamical magnitude range for which our survey is complete. In some cases, less luminous objects are seen (maximum dynamical range is 9\,mag). Observing dates, telescope, filters, exposure times, and the mean spatial resolution calculated as the average {\sc fwhm} of the images are all given in Table~\ref{photlog}.

We performed differential photometry and relative astrometry of all star-like sources that are found within a radius of 25\arcsec~from the low-metallicity source. For about 10\%~of our sample, we also investigated up to 1\farcm5. Based on the lower envelope delineated by the distance--metallicity relation of Fig$.$~\ref{dist}, and for an average seeing of 2\arcsec~(see Table~\ref{photlog}), we can resolve projected separations larger than 30\,AU for roughly solar metallicities, and larger than 200\,AU for [Fe/H]\,=\,$-$3. Instrumental magnitudes of both targets and companion candidates are determined by considering constant circular apertures that do not overlap with other sources. The magnitude difference of a candidate with respect to its metal-depleted star is computed by subtracting instrumental magnitudes. For bright sources, photometry is accurate to $\pm$0.05\,mag. The photometric uncertainty increases up to $\pm$0.25\,mag for the faintest sources.  Photometric diagrams will be discussed in Sect.~\ref{discussion}.

Projected separations and position angles are calculated from the centroids of the objects, which are obtained by fitting a gaussian function to the radial profile distribution of the sources. Common proper motion visual binaries are recognized because the members of a system should keep a rather constant separation and position angle within intervals of years, while changing their apparent location with respect to background sources. We note that the accuracy of the centroids of bright sources is typically $\pm$0.2\,pix, being about $\pm$1\,pix for objects close to the detection limit. This, in addition to the uncertainty of the {\sc ccd} plate scales, yield astrometric precisions between 0\farcs05 and 0\farcs4. The uncertainty in position angles is in the range 0.5--1\,deg, which also accounts for small offsets in the N-E orientation of the instruments. We summarize our photometric and astrometric results in Table~\ref{photlog}.

Target stars with nearby red companion candidates were observed at different epochs years apart, allowing us to identify common proper motion pairs. We also used red images provided by the Digitized Sky Surveys to complete our astrometric study. In addition, we searched the literature and {\sc simbad} for identifying wide binaries with large and very short separations, which we cannot resolve in our data. As a result, we found that G\,128-064 and G\,153-067 have proper motion companions located at 4\farcm3 (Allen et al$.$ \cite{allen00}) and 16\farcm6 (Gliese \& Jahreiss \cite{gliese88}), respectively, which obviously are off limits of the detectors used in our survey. We also found Hipparcos binaries (e.g., G\,230-049, G273-001) that have separations significantly shorter than the typical spatial resolution of our images. 

Of the 473 stars in the original sample, we detected 29 previously known binaries and multiple systems with separations between 0\farcs17 and 17\arcmin~(Luyten \cite{luyten79}; Perryman et al$.$ \cite{perryman97}; Salim \& Gould \cite{salim03}), and 13 new proper motion companions with separations between 2\arcsec~and 1\farcm9. Their photometric and astrometric data are listed in Tables~\ref{oldbin} (known pairs) and~\ref{newbin} (new pairs). In these tables, we provide the name of the companion, the metallicity and optical magnitudes of the primary star, the Hipparcos distance to the system (when available), the distance given by Carney et al$.$ (\cite{carney94}), and the projected separation in AU (the Hipparcos distance is preferred over the Carney et al$.$'s value). Optical magnitudes of primaries were taken from the Hipparcos database as well as from the literature (Stauffer \& Hartmann \cite{stauffer86}; Ryan \cite{ryan89}; Greenstein \cite{greenstein89}; Weis \cite{weis91}; Ryan \cite{ryan92}). 

Three of the new proper motion companions of Table~\ref{newbin} (G\,090-036\,B, G\,116-009\,B, and G\,176-046\,D) were previously reported by Mart\'\i n et al$.$ (\cite{martin95}). They are early discoveries of the survey described here. The optical low-resolution spectroscopy of G\,116-009\,B and G\,176-046\,D, presented in this paper and in Gizis \& Reid (\cite{gizis_reid97}) confirm them as low-metallicity late-type subdwarfs.

\begin{table*}
\caption{2MASS photometry of primaries and companions.\label{2massphot}}
\scriptsize
\begin{center}
\begin{tabular}{lrrr|lrrr}
\hline
\hline
Object        &
\multicolumn{1}{c}{$J$} &
\multicolumn{1}{c}{$H$} &
\multicolumn{1}{c|}{$K_s$}&
Object        &
\multicolumn{1}{c}{$J$} &
\multicolumn{1}{c}{$H$} &
\multicolumn{1}{c}{$K_s$}\\
\hline
G\,009-003\,A     &  9.676(.021) &  9.069(.021) &  8.843(.022) &  G\,172-016\,B     & 13.992(.041) & 13.408(.049) & 13.195(.042) \\                               
G\,009-003\,B     & 13.182(.021) & 12.584(.027) & 12.321(.023) &   G\,173-002        &  8.492(.019) &  8.048(.018) &  7.976(.023) \\                    
G\,009-047\,A     &  6.105(.034) &  5.594(.027) &  5.551(.021) &   G\,173-003        & 10.752(.018) & 10.174(.015) &  9.903(.020) \\                    
G\,009-047\,B     & 14.051(.028) & 13.366(.036) & 13.191(.030) &   G\,176-046\,A$^e$ &11.093(~~~~~) &10.608(~~~~~) & 10.496(.026) \\ 
G\,026-008\,A     &  8.877(.023) &  8.440(.055) &  8.352(.021) &   G\,176-046\,D$^e$ &12.276(~~~~~) &11.822(~~~~~) & 13.501(.049) \\ 
G\,026-008\,B     & 11.523(.040) & 11.022(.042) & 10.761(.039) &   G\,188-022\,A     &  8.984(.021) &  8.737(.038) &  8.645(.022) \\                    
G\,026-009        &  7.770(.018) &  7.240(.038) &  7.082(.029) &   G\,188-022\,B     & 13.471(.100) & 13.035(.107) & 12.859(.107) \\ 
G\,026-010        & 14.894(.086) & 15.050(.086) & 15.217(.149) &   G\,194-037\,A     & 13.582(.023) & 13.128(.030) & 13.044(.029) \\                    
G\,027-008\,A     & 10.213(.023) &  9.923(.026) &  9.870(.025) &   G\,194-037\,B     & 15.353(.050) & 14.840(.064) & 14.472(.075) \\                    
G\,027-008\,B     & 13.942(.056) & 13.516(.043) & 13.261(.051) &   G\,197-049        &  6.875(.027) &  6.245(.017) &  6.059(.017) \\                    
G\,028-040        &  9.886(.019) &  9.468(.030) &  9.407(.021) &   G\,197-050        &  9.171(.022) &  8.659(.021) &  8.387(.015) \\                    
LP\,581-80        & 11.727(.019) & 11.095(.031) & 10.922(.018) &   G\,204-049\,A$^f$ &  9.426(.023) &  9.034(.026) &  8.903(.020) \\                    
G\,032-046\,A     &  8.335(.029) &  7.928(.051) &  7.877(.031) &   G\,204-049\,B$^a$ &        ---   &         ---  &         ---  \\                    
G\,032-046\,B$^a$ &        ---   &         ---  &        ---   &   G\,214-001\,A     & 10.756(.024) & 10.443(.030) & 10.393(.018) \\                    
G\,059-032\,A     &  7.401(.034) &  6.953(.067) &  6.847(.024) &   G\,214-001\,B     & 14.511(.100) & 14.077(.089) & 13.945(.080) \\                    
G\,059-032\,B     & 14.106(.027) & 13.723(.042) & 13.215(.033) &   G\,216-045\,A     &  9.855(.023) &  9.579(.021) &  9.496(.019) \\                    
G\,061-024        &  7.681(.018) &  7.367(.015) &  7.302(.017) &   G\,216-045\,B     & 15.019(.045) & 14.454(.049) & 14.138(.051) \\                    
G\,061-023        &  8.481(.024) &  8.223(.027) &  8.120(.023) &   G\,217-004\,A     &  9.946(.020) &  9.569(.018) &  9.474(.016) \\                    
G\,063-036$^b$    &  7.643(.023) &  7.067(.026) &  6.828(.021) &   G\,217-004\,B     & 14.191(.032) & 13.619(.037) & 13.397(.038) \\                    
G\,066-22\,A      &  9.030(.023) &  8.577(.023) &  8.634(.027) &   G\,230-047\,A     &  8.735(.035) &  8.360(.026) &  8.297(.021) \\                    
BD+06\,2932\,B    &  9.304(.095) &  8.858(.049) &  8.816(.057) &   G\,230-047\,B     & 11.745(.024) & 11.196(.019) & 10.959(.024) \\                    
G\,069-004        &  7.035(.018) &  6.659(.023) &  6.602(.016) &   G\,230-049\,AB$^c$&  7.263(.021) &  6.916(.018) &  6.866(.020) \\                    
G\,069-004\,N     &  7.790(.020) &  7.466(.023) &  7.397(.031) &   G\,242-074\,A     &  8.693(.029) &  8.404(.044) &  8.360(.021) \\                    
G\,090-036\,AB$^c$& 11.382(.021) & 11.046(.022) & 10.962(.020) &   G\,242-074\,B     & 12.214(.025) & 11.704(.031) & 11.478(.023) \\                    
G\,093-027\,A     & 10.120(.041) &  9.610(.046) &  9.505(.035) &   G\,251-054\,A     &  8.711(.039) &  8.333(.038) &  8.261(.026) \\                    
G\,093-027\,B$^d$ & 11.467(.038) & 10.834(.032) &  9.463(.055) &   G\,251-054\,B     & 12.541(.029) & 12.027(.035) & 11.888(.023) \\                    
G\,095-057\,A     &  6.618(.026) &  6.192(.021) &  6.120(.023) &   G\,252-049\,AB$^c$&  6.888(.021) &  6.242(.020) &  6.084(.020) \\                    
G\,095-057\,B     &  6.949(.023) &  6.481(.024) &  6.347(.023) &   G\,255-038\,A     &  7.799(.037) &  7.251(.044) &  7.108(.024) \\                    
G\,116-009\,A     & 12.660(.021) & 12.161(.021) & 12.084(.021) &   G\,255-038\,B     &  9.574(.022) &  8.942(.030) &  8.709(.022) \\                    
G\,116-009\,B     & 15.331(.045) & 14.710(.054) & 14.571(.090) &   G\,262-021        & 12.264(.042) & 11.769(.044) & 11.743(.037) \\                    
G\,128-064\,A     &  8.251(.026) &  7.881(.015) &  7.836(.019) &   G\,262-022        & 12.240(.029) & 11.638(.033) & 11.495(.024) \\                    
G\,128-064\,B     & 10.222(.028) &  9.763(.026) &  9.668(.023) &   G\,266-029\,AB$^c$&  8.263(.018) &  8.095(.020) &  8.078(.027) \\                    
G\,128-077\,A     & 11.655(.021) & 11.154(.024) & 11.068(.020) &   G\,273-001        &  7.082(.019) &  6.860(.020) &  6.771(.023) \\                    
G\,128-077\,B     & 13.531(.024) & 12.960(.026) & 12.828(.031) &   LHS\,541          & 13.032(.027) & 12.559(.028) & 12.414(.026) \\                    
G\,153-067        &  8.055(.024) &  7.637(.040) &  7.517(.024) &   G\,273-152\,A$^g$ &  8.694(.026) &  8.291(.029) &  8.227(.042) \\                    
G\,017-027        & 11.086(.022) & 10.541(.022) & 10.319(.019) &   G\,273-152\,B$^a$ &        ---   &          --- &         ---  \\                    
G\,171-050\,A     & 11.939(.023) & 11.640(.022) & 11.603(.023) &   BD+16\,2116\,AB$^c$& 5.894(.021) &  5.550(.016) &  5.467(.021) \\                    
G\,171-050\,B     & 15.909(.060) & 15.452(.087) & 15.191(.126) &   BD+23\,2207\,A    &  4.995(.260) &  4.696(.228) &  4.512(.017) \\                    
G\,172-016\,A     &  9.748(.023) &  9.420(.020) &  9.333(.018) &   BD+23\,2207\,B    &  8.355(.026) &  7.794(.047) &  7.593(.026) \\                    
\hline
\end{tabular}
\end{center}
$^a$~No 2MASS photometry available.\\
$^b$~Combined photometry of G\,063-036 and GJ\,516\,B.\\
$^c$~Combined photometry.\\
$^d$~Error in $K_s$: the detection was not geometrically possible.\\
$^e$~$JH$ are upper limits to the photometry. $K_s$ is derived from a profile-fitting technique.\\
$^f$~Magnitudes are derived from a profile fitting measurement. Not likely contaminated by the companion.\\
$^g$~Photometry is partially contaminated by the companion.\\
\end{table*}

\subsection{2MASS near-infrared photometry}
Table~\ref{2massphot} provides the $JHK_s$ magnitudes of both the primary and secondary stars, which we obtained from the 2MASS all-sky survey data archive. All the sources are unambigously identified. From our list, companions separated by less than 4\arcsec~are not well spatially resolved by 2MASS. In these systems, except for G\,204-049, photometry is contaminated by the presence of the companion (see the notes to the Table). We remark that the 2MASS $J$ and $H$ magnitudes of G\,176-046\,A and G\,176-046\,D are only the upper limits to the real values; on the contrary, the $K_s$ data are reliable because the two objects were resolved by 2MASS and the photometry was obtained by a profile-fitting technique. Because of contaminated or unresolved near-infrared photometry, the following new pairs are not included in the near-infrared plots of Sect.~\ref{discussion}: G\,090-036, G\,093-027, G\,176-046, G\,204-049, and G\,273-152.

\subsection{Optical spectroscopy}
We obtained optical low-resolution spectroscopy of eight new proper motion companions listed in Table~\ref{speclog}. Six of them were observed using the double-arm ISIS spectrograph and the 4.2\,m William Herschel Telescope (WHT) at the ORM. For comparison purposes, we also observed BD\,+06\,2932\,B (companion to G\,066-022) and LHS\,541 (vB\,12), the low-metallicity dwarf companion to LHS\,540 (G\,273-1), which was discovered by van Briesbroeck (\cite{vanbries61}). On 2002 Aug 23 we used the 570\,nm dichroic, the R600R grism in the red arm, and the R300B grism in the blue arm. The pixel size of the {\sc eev ccd} detectors and the binning of a factor of 2 along the spectral direction in the red arm provided nominal dispersions of 0.865\,\AA/pix (blue) and 0.889\,\AA/pix (red). On 2003 Sep 11 we only used the red arm, the R158R grism, the order-blocking filter GG495, and the same detector as in the previous run. This instrumental setup provided a lower nominal dispersion of 1.63\,\AA/pix (no binning was applied). Older data were taken on 1994 Jul 1--2, using the R158B grating in the blue arm and the R158R grating in the red arm, which provided nominal dispersions of 2.9 and 2.7\,\AA/pix, respectively. We always observed through a slit-width of 1\arcsec; all data were collected at parallactic angle to minimize light losses. The seeing in the visible was rather poor (2\arcsec) on 23 Aug 2002 and on Jul 1994, and 1\arcsec~on 11 Sep 2003. This latter night was hampered by clouds and was clearly non-photometric. 

Optical spectra of two slightly metal-poor, new companions were obtained with the DOLORES spectrograph and the 3.5\,m Telescopio Nazionale Galileo (TNG) at the ORM on 2003 Dec 12. We used a slit-width of 1\farcs5, the low-resolution grism LR-R and a 2048\,$\times$\,2048 LORAL detector, which provided a nominal dispersion of 2.88\,\AA/pix and a final resolution of 13\,\AA. These observations were not carried out at parallactic angle.  

\begin{table}
\caption{Log of spectroscopic observations.\label{speclog}}
\begin{center}
\begin{tabular}{lcccc}
\hline
\hline
Object         & Obs$.$ Date & Exp.         &$\Delta \lambda$& Coverage  \\
               &             &  (s)         &(\AA)           &   (nm)    \\
\hline			                                              
\multicolumn{5}{c}{WHT spectra}                                          \\
BD+06\,2932\,B &  2 Jul 1994 & 300          & 6.1            & 386--590  \\
               &  2 Jul 1994 & 300          & 5.7            & 605--900  \\
G\,116-009\,B  &  1 Jul 1994 & 300,600      & 6.1            & 386--590  \\
               &  1 Jul 1994 & 300          & 5.7            & 605--900  \\
G\,128-077\,B  & 23 Aug 2002 & 2$\times$400 & 3.3            & 391--580  \\
               & 23 Aug 2002 & 2$\times$400 & 3.4            & 641--775  \\
G\,172-016\,B  & 11 Sep 2003 & 2$\times$600 & 6.2            & 575--1000 \\
G\,176-046\,D  &  2 Jul 1994 & 2$\times$600 & 6.1            & 386--590  \\
               &  2 Jul 1994 & 2$\times$600 & 5.7            & 605--900  \\
G\,188-022\,B  & 11 Sep 2003 & 3$\times$600 & 6.2            & 575--1000 \\
G\,214-001\,B  & 11 Sep 2003 & 3$\times$600 & 6.2            & 575--1000 \\
LHS\,541       & 23 Aug 2002 & 3$\times$300 & 3.3            & 391--580  \\
               & 23 Aug 2002 & 3$\times$300 & 3.4            & 641--775  \\
               & 11 Sep 2003 & 600          & 6.2            & 575--1000 \\   
\hline			                                              
\multicolumn{5}{c}{TNG spectra}                                          \\
G\,216-045\,B  & 12 Dec 2003 & 2$\times$900 & 13             & 525--900  \\
G\,273-152\,B  & 12 Dec 2003 & 60           & 13             & 525--900  \\
\hline
\end{tabular}
\end{center}
\end{table}

Raw frames were bias-subtracted and flat-field corrected using standard techniques. Spectra were optimally extracted with the {\sc twodspec} package of {\sc iraf}. The error in wavelength calibration, which was performed with atomic lines of Cu, Ar and Ne, is around 2.5--5\%. Instrumental response was removed to an accuracy of 10\%~using observations of the spectroscopic standard stars G\,191-B2B, BD+25\,3941, BD+28\,4211 and Hilter\,102, which have fluxes available in the {\sc iraf} database. The log of the spectroscopic observations appears in Table~\ref{speclog}, where we include exposure time per object, final spectral resolution, and spectral coverage. 

\begin{figure}
\centering
\includegraphics[width=8.5cm]{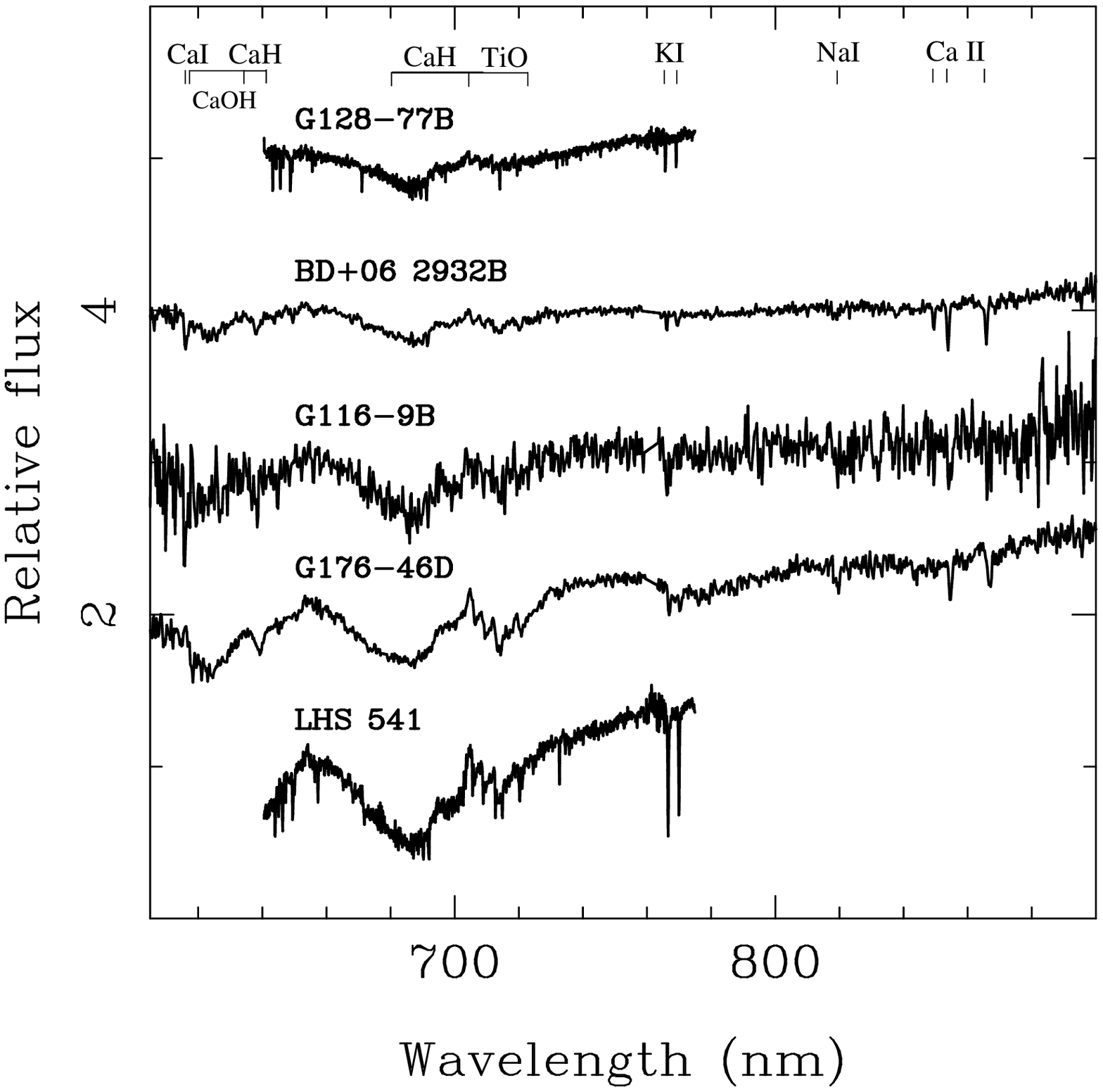}
\includegraphics[width=8.5cm]{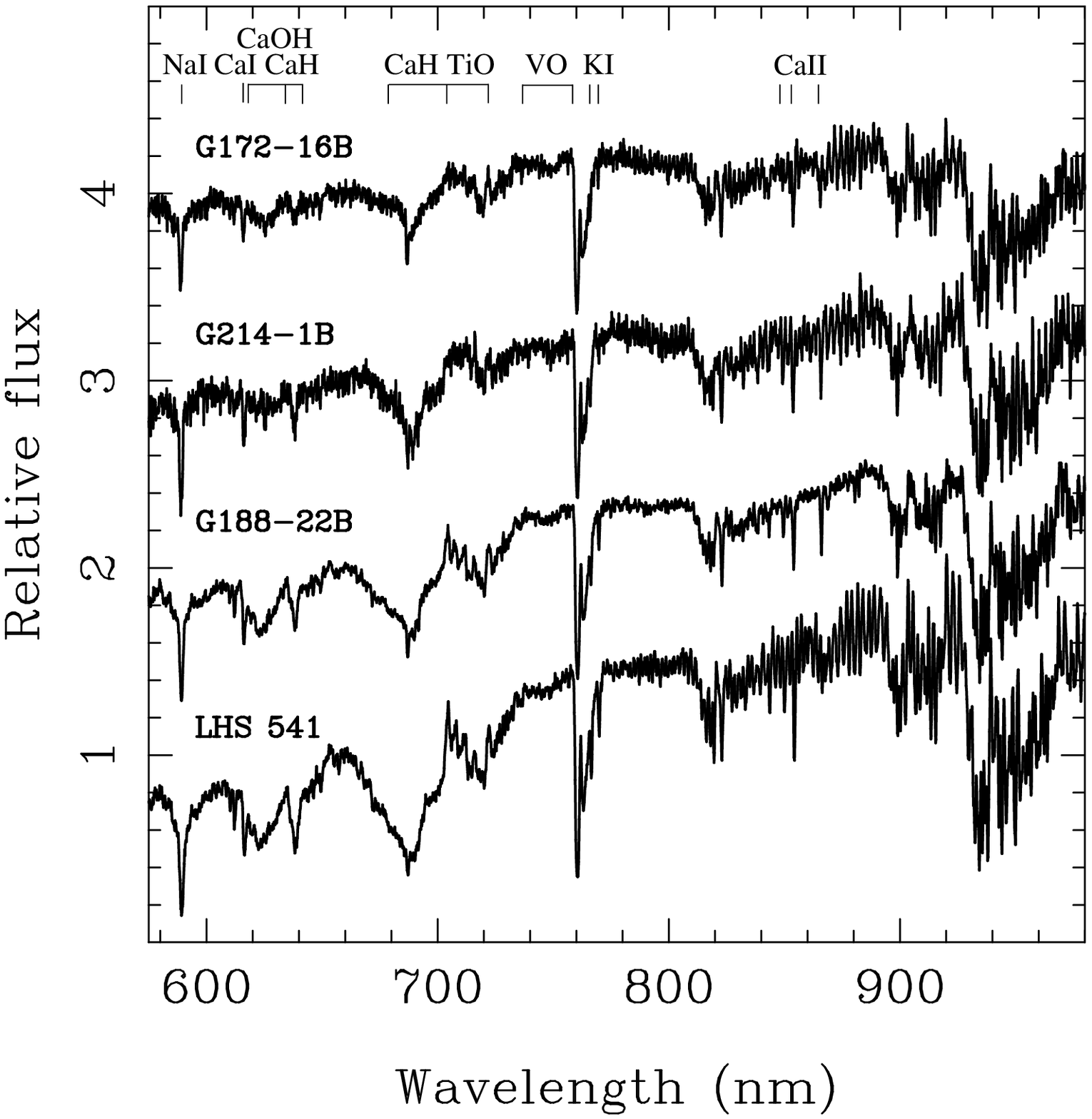}
\caption{WHT spectra of subdwarfs ordered by decreasing spectral type from top to bottom. Data are normalized to unity at 660--662.5\,nm, and are displaced upwards by 1 for clarity. The top panel displays spectra corrected for telluric absorptions. The bottom panel shows spectra with strong telluric features around 687, 718, 760, 815, 900, and 932\,nm. Fringing is observed redwards of 800\,nm. Some spectral features are identified at the top of each panel.}
\label{redspec} 
\end{figure}

\begin{figure}
\centering
\includegraphics[width=8.5cm]{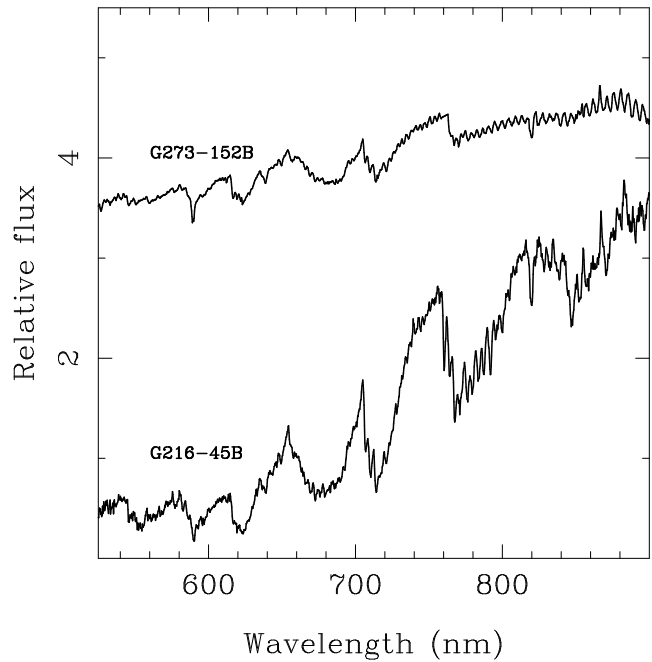}
\caption{TNG spectra of slightly metal-poor, M-type proper motion companions. Data are normalized to unity at 660--662.5\,nm, and are displaced upwards by 3 for clarity. Note that these spectra were not taken at parallactic angle; hence, the flux ratio between very red and blue wavelengths may be incorrect. Telluric absorption and fringing are removed.}
\label{tngspectra} 
\end{figure}

\begin{figure}
\centering
\includegraphics[width=8.5cm]{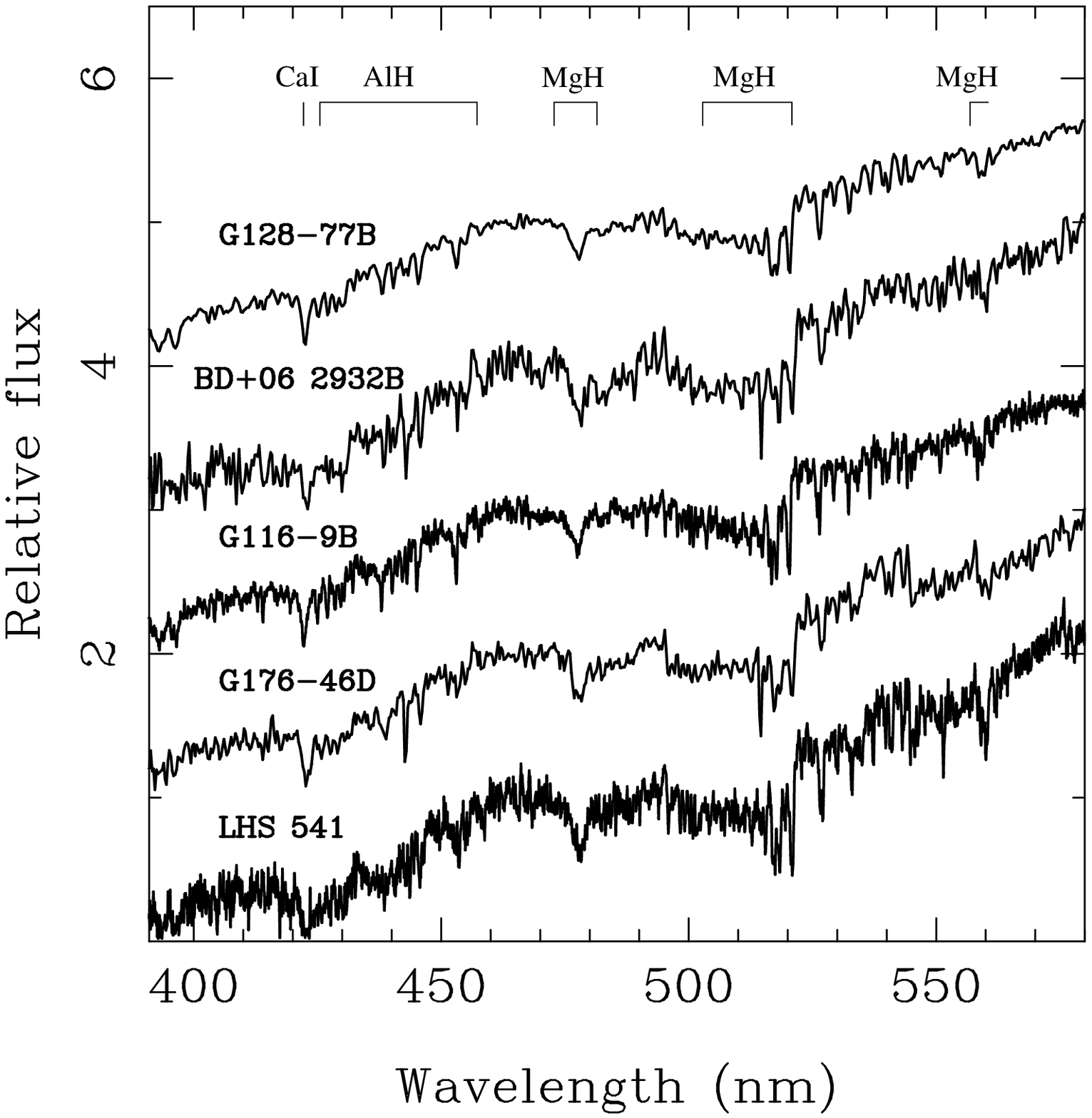}
\caption{WHT blue spectra normalized to unity at 461--463\,nm, and offset upwards by 1 for clarity. Spectral features are identified following Ake \& Greenstein (\cite{ake80}) and Dahn et al$.$ (\cite{dahn95}).}
\label{bluespec} 
\end{figure}

The resulting spectra are depicted in Figs.~\ref{redspec}, \ref{tngspectra}, and~\ref{bluespec}. The 2003 WHT and TNG spectra are affected by a strong fringing pattern redwards of 800\,nm. We did not attempt to remove telluric features from the WHT data because of the very fast changes of the sky conditions during the observations. On the contrary, the TNG data were reasonably corrected for telluric absorption and fringing, although some fringing residuals are observed in the spectrum of G\,273-152\,B (Fig.~\ref{tngspectra}). The 1994 and 2002 red data were also corrected for telluric absorption. This was accomplished by observing G-type stars at similar airmass shortly before the target observation and interpolating the continuum across the telluric bands. This correction leads to a better detection of the K\,{\sc i} atomic doublet at 766.5 and 769.9\,nm, as shown in the upper panel of Fig.~\ref{redspec}.

The spectra show enhanced absorption features of the hydrides MgH and CaH, labeled in Figs.~\ref{redspec} and~\ref{bluespec}, which are typical of cool, low-metallicity subdwarfs. Regarding oxides, the data have rather weak (as compared to solar-metallicity stars of similar types) but detectable TiO absorption at around 710\,nm. Also noticeable are the atomic features ascribed to Na\,{\sc i}, K\,{\sc i} and Ca\,{\sc ii}. We checked for the presence of the H$\alpha$ line. At the resolution of our spectra, H$\alpha$ is not seen in emission in any source. Hence, there is no evidence of chromospheric activity, as expected for old, metal-poor stars. Additionally, this suggests that none of them are interacting close binaries. 

\subsubsection{Spectral typing}
We followed the classification scheme developed by Reid et al$.$ (\cite{reid95}), Gizis (\cite{gizis97}) and L\'epine et al$.$ (\cite{lepine03a}) to obtain the spectral type of our objects. This scheme is based on the molecular indices TiO5, CaH2 and CaH3, which are excellent indicators of temperature and metallicity. The CaH2 and CaH3 indices measure the strength of the CaH bands blueward of 705\,nm, and the TiO5 index measures the TiO band head redward of 705\,nm. Gizis (\cite{gizis97}) claimed that the CaH3 index is particularly well suited for the ``numerical'' classification of metal-poor and extreme metal-poor dwarfs in the range K7--M6. Classes are K7, M0, M1 and so on. The measured indices for our stars are given in Table~\ref{specdata}; the uncertainty is about 5--10\%. We also provide the final averaged spectral types obtained to an accuracy of $\pm$\,0.5 subclasses. We note, however, that the spectral classes of G\,216-045\,B and G\,273-152\,B may suffer from larger error bars, i.e., one subclass, because they were not observed at parallactic angle. We adopted the nomenclature widely used in the literature, and spectral types have the prefix ``sd'' or ``esd'' that stand for ``subdwarf'' or ``extreme subdwarf'', respectively. All of our WHT spectroscopically observed objects are extreme subdwarfs according to their very low-metallicities. The TNG stars are classified as ``sdM'' because of their relatively high metallicity. Our spectral measurements of the objects in common with Gizis \& Reid (\cite{gizis97}), i.e., LHS\,541, G\,116-009\,B, and G\,176-046D, are fully consistent with previous published values. The new companions have spectral classes ranging from very late-K to early-M, which is in agreement with their optical and near-infrared colors.

\begin{table}
\caption{Spectroscopic data.\label{specdata}}
\begin{center}
\begin{tabular}{lcccc}
\hline
\hline
Object         & TiO5  & CaH2  & CaH3  & SpT          \\
               &       &       &       & ($\pm$\,0.5) \\
\hline		
BD+06\,2932\,B & 0.872 & 0.819 & 0.900 & esdK7.3 \\
G\,116-009\,B  & 0.9:  & 0.726 & 0.869 & esdM0.0 \\
G\,128-077\,B  & 0.928 & 0.823 & 0.913 & esdK7.2 \\
G\,172-016\,B  & 0.902 & 0.818 & 0.860 & esdK7.5 \\
G\,176-046\,D  & 0.709 & 0.622 & 0.809 & esdM1.0 \\
G\,188-022\,B  & 0.813 & 0.621 & 0.764 & esdM1.0 \\
G\,214-001\,B  & 1.0:  & 0.765 & 0.829 & esdM0.0 \\
G\,216-045\,B  & 0.526 & 0.42: & 0.713 & sdM3.0  \\
G\,273-152\,B  & 0.717 & 0.656 & 0.838 & sdM1.0  \\
LHS\,541       & 0.731 & 0.469 & 0.641 & esdM3.0 \\
\hline
\end{tabular}
\end{center}
{\sc notes.--} A semicolon indicates error bars twice larger than those given in the text. The uncertainty in the spectral classification of G\,216-045\,B and G\,273-152\,B may be one subclass.
\end{table}

The spectral types of G\,172-016\,B and G\,214-001\,B are quite alike within error bars. However, a detailed inspection of the red spectra of Fig.~\ref{redspec} reveals that the intensity of the molecular bands at around 620 and 680\,nm is different:  G\,214-001\,B shows less molecular absorption. This is explained by the more metal-depleted atmosphere of this object. Because they have similar metallicity ([Fe/H]\,$\sim$\,$-$1.5), G\,172-016\,B, G\,116-009\,B, G\,188-022\,B and LHS\,541 are better examples of the noticeable effect of decreasing effective temperature in metal-deficient dwarfs: the Na\,{\sc i} and K\,{\sc i} resonance doublets, and all molecular absorptions become stronger at lower temperatures. However, and despite the fact that the objects shown in Fig.~\ref{bluespec} have quite different classes, their blue spectroscopic data appear rather similar, except for the slightly larger slope towards red wavelengths of LHS\,541, which is the latest type subdwarf. We note that red spectra (550--800\,nm) are preferred over blue spectra (400--550\,nm) for the spectral classification of low-metallicity stars, since the red spectroscopic features are more sensitive to temperature changes than the blue ones.

\section{Discussion \label{discussion}}

From our survey, we found 13 new common proper motion pairs, and retrieved a total of 29 known pairs. This suggests that about one third of the stellar wide companions are missing in previous proper motion searches. 

\subsection{Photometric diagrams}

The optical and near-infrared photometry of Tables~\ref{oldbin}, \ref{newbin} and~\ref{2massphot} suggests that most of the proper motion companions are dwarfs. There are two exceptions: G\,026-010, which is a known white dwarf companion to the K3-type star G\,026-009 (Wegner \cite{wegner73}), and G\,032-046\,B, which is companion to a G8-type star (Allen et al$.$ \cite{allen00}). G\,026-010 shows quite blue optical and 2MASS colors that are typical of normal, hot white dwarfs. However, G\,032-046\,B appears redder, with M($V$)\,=\,14.30\,$\pm$\,0.15\,mag and ($V-I$)\,=\,0.84\,$\pm$\,0.15\,mag. Despite its relatively brightness in the visible, it is not detected by 2MASS ($J$\,$\ge$\,17), which indicates that G\,032-046\,B is indeed a ``blue'' object ($I-J$\,$\le$\,0.8). In fact, it is listed in the catalog of white dwarfs by Luyten (\cite{luyten70}). In a M($V$) vs$.$ ($V-I$) diagram like that of Fig$.$~10 of Monet et al$.$ (\cite{monet92}), G\,032-046\,B lies at the bottom of the degenerate sequence. Hence, this object is a cool, low-metallicity white dwarf. 

\begin{figure}
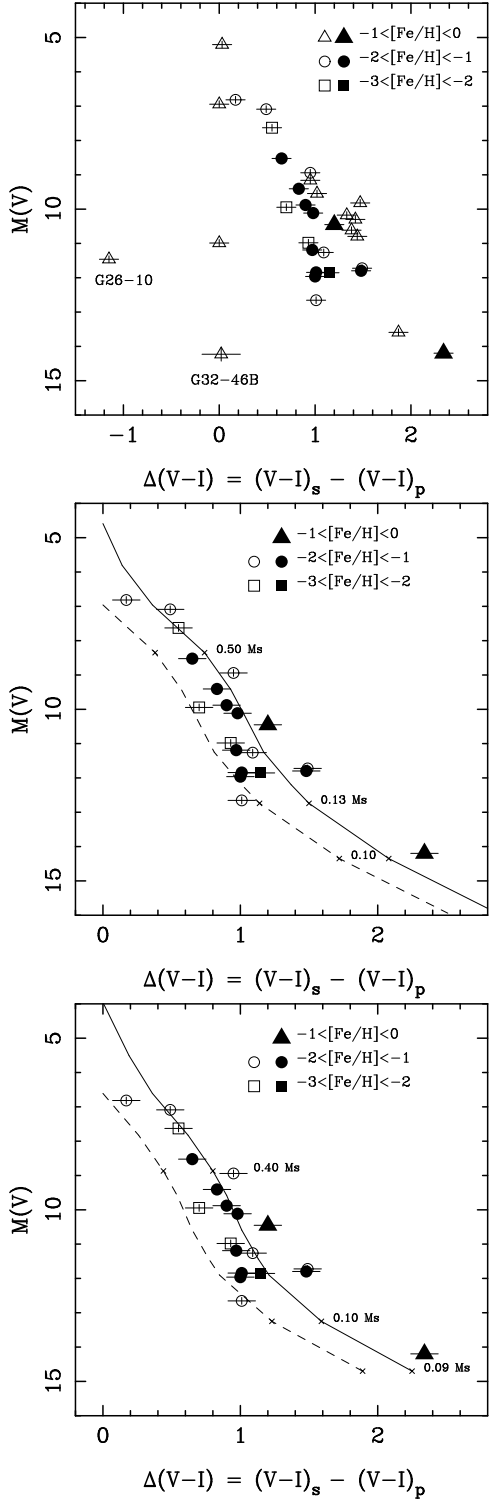

\centering
\includegraphics[width=6.4cm]{phot1.ps}
\includegraphics[width=6.4cm]{phot2.ps}
\includegraphics[width=6.4cm]{phot3.ps}
\caption{{\sl Top panel:} Optical color-magnitude diagram of proper motion companions. Different symbols denote different intervals of metallicity (in logarithmic scale), as indicated within the figure. Filled symbols stand for the new discoveries, and open symbols represent previously known secondaries. {\sl Middle panel:} model calculations by Baraffe et al$.$ (\cite{baraffe97}) for metallicity $-$1\,dex are overplotted. The track with $(V-I)$ color difference computed for a primary mass of 0.8\,$M_{\odot}$ is depicted with a full line, and that computed for a primary mass of 0.6\,$M_{\odot}$ is displayed with a dashed line. {\sl Bottom panel:} as in the previous panel, but for a metallicity of $-$2\,dex. }
\label{hr} 
\end{figure}

Figure~\ref{hr} depicts the optical color-magnitude diagram of the proper motion companions with $VI$ photometry. The two white dwarfs are labelled. To transform observed magnitudes into absolute magnitudes we used the distance to the primary stars from the Hipparcos database. If no Hipparcos data were available, we used the values given by Carney et al$.$ (\cite{carney94}) and Laird et al$.$ (\cite{laird88}). We assumed that the distance to the companions is equal to the distance to their primaries. In the Figure, we plotted different symbols for different metallicities. Because no calibrated photometry in the $V$, $R$ and $I$ bands is available for all of the objects, the x-axis of Fig.~\ref{hr} represents the $(V-I)$ color difference [$\Delta (V-I)$] between the companion and its primary star. We note that this quantity is reddening-free. As expected, white dwarfs delineate a degenerate sequence at low luminosities, and dwarfs extend to very red colors. Dwarf companions that are fainter than their primary stars are also redder. 

\begin{figure}
\centering
\includegraphics[width=8.5cm]{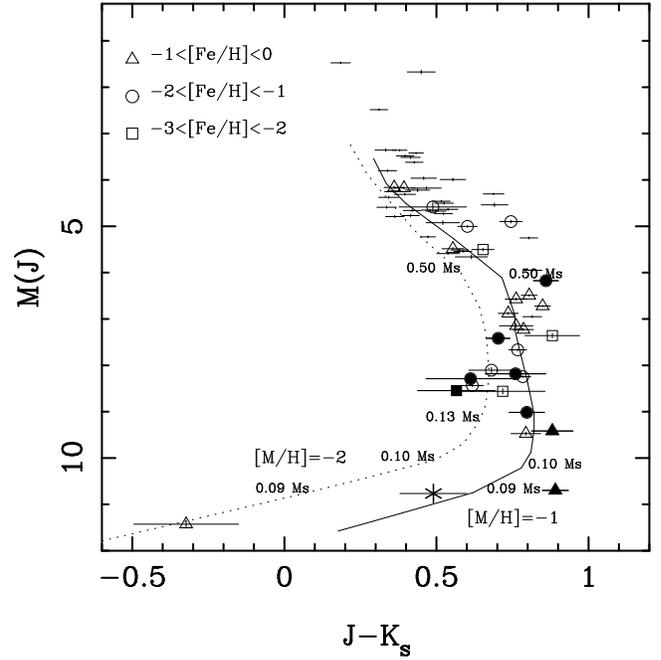}
\caption{2MASS near-infrared color-magnitude diagram for the wide binaries in our sample and the esdM8.0 star LSR\,1425+7102 (asterisk). Primary stars are plotted as error bars without any symbol. New companions are plotted with filled symbols. The object with the bluest color is a white dwarf. Overplotted onto the data are the 10-Myr isochrones from Baraffe et al$.$ (\cite{baraffe97}). The faint end of the isochrones indicates the location of the substellar limit for each metallicity.}
\label{jhk} 
\end{figure}

\begin{figure}
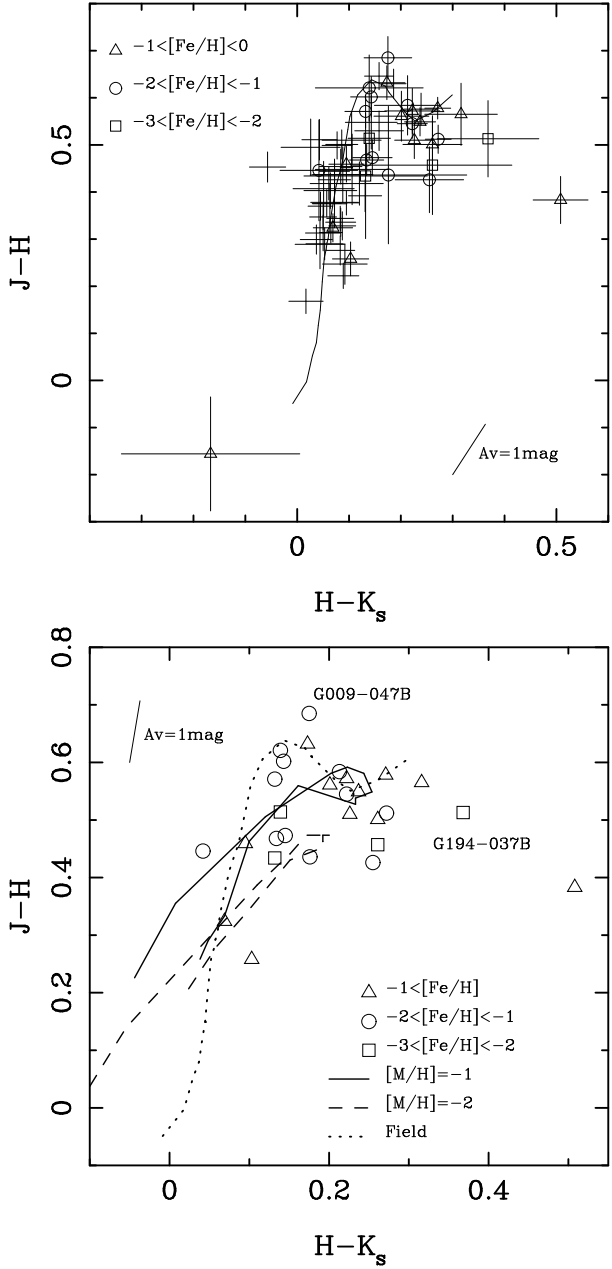

\centering
\includegraphics[width=8cm]{jhkcolor.ps}
\includegraphics[width=8cm]{jhkcolor_detail.ps}
\caption{{\sl Top panel:} 2MASS color-color diagram for the wide binaries in our sample. Both new and known companions are plotted with open symbols. Primary stars are plotted as error bars without any symbol. Overplotted onto the data is the field dwarf sequence from Bessell \& Brett (\cite{bessell88}). The object with the bluest color is a white dwarf. {\sl Bottom panel:} Detail of the color-color diagram where models (solid and dashed lines) calculated for two different metallicities are overplotted (Baraffe et al$.$ \cite{baraffe97}). Error bars are omitted for clarity. The reddening vector corresponding to A($V$)\,=\,1\,mag (Rieke \& Lebofski \cite{rieke85}) is plotted in every panel.}
\label{jhkcolor} 
\end{figure}

The observed scatter in the dwarf sequence of Fig.~\ref{hr} is very likely due to the different colors of the primaries (which have an effect on the $\Delta (V-I)$ values), unresolved binarity, photometric errors, uncertainties in the distance determination, and different degrees of atmospheric metallic composition. Despite all these uncertainties, it is obvious that the most metal-rich stars appear brighter for a given color. This is due to the fact that most of the primary stars have similar optical colors (spectral types F8--K0), and that metal-depleted stars are more subluminous for lower metallicities. Besides the two white dwarfs, there are two outliers in Fig.~\ref{hr}: GJ\,516\,B and G\,197-50, which appear very faint for their colors. Their primary stars are significantly much redder (spectral type M) than other primaries.

To compare the location of the observed photometric sequence of metal-poor, dwarf companions to theoretical models, we overplotted 10-Gyr isochrones from Baraffe et al$.$ (\cite{baraffe97}) in the middle and bottom panels of Fig.~\ref{hr}. Here, we did not display known secondaries with [Fe/H]\,$>$\,$-1$ for clarity. This has no impact in the following discussion. The metallicities of the theoretical tracks depicted in the Figure are $-1$ and $-2$\,dex; the majority of the new proper motion companions have a metal content within this range. According to Carney et al$.$ (\cite{carney94}), the mass of all of the primary stars spans from 0.55 up to 0.8\,$M_{\odot}$, the mean value being 0.67\,$M_{\odot}$. We note, however, that the most metal-rich primaries have larger masses than the stars with less metal abundance. To compute the predicted $\Delta (V-I)$ color differences, we considered masses of 0.6 and 0.8\,$M_{\odot}$, i.e., two stars of the same mass and metallicity have $\Delta (V-I)$\,=\,0.0\,mag, and the color difference increases for smaller mass ratios. As seen from Fig.~\ref{hr}, the models nicely reproduce the trend described by the observations.   Baraffe et al.'s models also reproduce reasonably well the entire main sequence of globular clusters.

We also produced the 2MASS near-infrared color-magnitude diagram shown in Fig$.$~\ref{jhk}. As in the previous figure, the 10-Myr isochrones from Baraffe et al$.$~(\cite{baraffe97}) are overplotted.  These authors provide near-infrared colors in the CIT photometric system. To convert them into the 2MASS reference system, we used the transformation equations given in Carpenter (\cite{carpenter01}). As in the visible, the agreement between the trend delineated by near-infrared observations and theory is noticeable. 

The near-infrared two-color diagram is depicted in Fig$.$~\ref{jhkcolor}. In the upper panel, overplotted is the sequence of field, solar-metallicity stars from Bessell \& Brett (\cite{bessell88}). We adopted the color transformations of Alonso et al$.$ (\cite{alonso94}) to convert Bessell \& Brett's data into the TCS system, which is very similar to the 2MASS photometric system (see B\'ejar et al$.$ \cite{bejar03}). The lower panel of Fig$.$~\ref{jhkcolor} displays an enlargement of the color-color diagram, where the two Baraffe et al.'s isochrones are considered and error bars are avoided for the sake of clarity. From this picture, metallicity effects become apparent: the ($J-H$) colors of the most metal-deficient companions (squares) appear blueshifted as compared to the colors of the most metal-rich stars (triangles), indicating that metallicity has a clear effect on the near-infrared colors. This signature is qualitatively reproduced by the models. Further discussion on the colors of subdwarfs and dwarfs is provided in Leggett (\cite{leggett92}) and Tinney et al$.$ (\cite{tinney93}). It is also remarkable the theoretically predicted turn-over to the blue (both $J-H$ and $H-K$), which is due to the increasing collision-induced absorption of molecular hydrogen with increasing density and decreasing temperature (Saumon et al$.$ \cite{saumon94}). This turn-over takes place at around 0.2\,$M_{\odot}$ and $T_{\rm eff}$\,=\,3600--4000\,K for the two metal-depleted abundances. A similar feature is observed in solar-metallicity dwarfs (e.g., Leggett et al$.$ \cite{leggett02}, and references therein), but it occurs at much cooler temperatures ($T_{\rm eff}$\,$\sim$\,1500\,K), within the substellar mass domain, and is explained by the appearance of strong methane absorptions at 1.6 and 2.2\,$\mu$m.  

Three proper motion companions show quite redder near-infrared colors than expected for their metallicity and brightness: G\,262-022 ([Fe/H]\,=\,$-1.07$), G\,009-047\,B ([Fe/H]\,=\,$-1.93$), and G\,194-037\,B ([Fe/H]\,=\,$-2.03$). Except for G\,059-032\,B, G\,009-047\,B and G\,194-037\,B have the reddest ($J-H$) and ($H-K_s$) colors, respectively, in our final list of companions (see Figs.~\ref{hr} and~\ref{jhk}). These objects appear shifted to red colors by more than 2-$\sigma$ their photometric uncertainties provided their atmospheric chemical composition is accurately determined. The excesses observed in G\,262-022 and G\,009-047\,B could be explained by small amounts of interstellar reddening (A$_V$\,$\sim$\,0.3--1\,mag), as illustrated by the reddening vector in the two panels of Fig.~\ref{jhk}. A fainter companion might also be contributing to the very red ($J-K_s$) color of G\,262-022. Carney et al$.$ (\cite{carney94}) detected some extinction in G\,262-021 (the primary star), but no extinction has been photometrically measured in G\,009-047\,A (Laird et al$.$ \cite{laird88}). The $K_s$-band excess of G\,194-037\,B is not compatible with interestellar reddening alone (no extinction is measured by Carney et al$.$ \cite{carney94}). From the models, less massive, unresolved companions are not expected to contribute significantly to the near-infrared flux excess below M($J$)\,=\,6\,mag, since the theoretical photometric sequence goes down to M($J$)\,=\,10\,mag almost vertically in the $J$ vs$.$ ($J-K_s$) diagram of Fig.~\ref{hr}. There are two possibilities: either there is a warm dusty disk surrounding G\,194-037\,B, or its low-metallicity is overestimated (most likely). We note that the very recent work by Latham et al$.$ (\cite{latham02}) gives [Fe/H]\,=\,$-1.0$ for G\,194-037\,A (the primary star). This is more in agreement with the location of the proper motion companion in Fig.~\ref{jhk}.

\subsection{Mass estimate}

An estimate of the mass of the proper motion secondaries can be obtained from the comparison of the objects' locii in color-magnitude diagrams with evolutionary tracks. As inferred from Figs.~\ref{hr} and~\ref{jhk}, the new companions have stellar masses between 0.1 and 0.5\,$M_{\odot}$, as do many of the previously known companions of our survey and field metal-depleted stars (Monet et al$.$ \cite{monet92}). 

One of the reddest, least massive, low-metallicity stars detected in our work is G\,216-045\,B. The companion G\,059-032\,B is actually the reddest one (mid-M spectral class according to its near-infrared colors), but its metallicity (assumed to be equal to that of the primary star) is very close to solar. According to Figs.~\ref{hr} and~\ref{jhk}, the mass of G\,216-045\,B can be estimated at 0.1--0.13\,$M_{\odot}$ if we adopt an age typical of the old disk. The spectral type of G\,216-045\,B is sdM3 on the basis of its optical spectrum. This star actually belongs to a multiple system formed by at least three members. The primary star, G\,216-045\,A (spectral type K1, Lee \cite{lee84}), is found to be a 1153\,day-period spectroscopic binary by Latham et al$.$ (\cite{latham02}). The projected semi-major axis of the spectroscopic double is 112.8\,AU, while G\,216-045\,B is separated by about 4150\,AU (see Table~\ref{newbin}) from the more massive stars. Multiple systems with wide companions are not rare among metal-deficient stars, as discussed in Mart\'\i n et al$.$ (\cite{martin95}).

Some later M-type subdwarfs have been recently identified in the field by means of proper motion measurements. L\'epine et al$.$ (\cite{lepine03b}) reported on the finding of a low-metallicity star with spectral type sdM8.0: LSR\,1425+7102. For comparison purposes, we plot its 2MASS photometry as an asterisk symbol in Fig.~\ref{jhk}. Its location in the Figure suggests that the atmospheric metal abundance of this star is around [M/H]\,=\,$-1$ (i.e., all metals are depleted by a factor of 10 as compared to the Sun). From the comparison with model calculations, its mass is likely between 0.085 and 0.09\,$M_{\odot}$, rather close to the substellar borderline for this metallicity (i.e, 0.083\,$M_{\odot}$, Baraffe et al$.$ \cite{baraffe97}). We do not detect in our survey objects as cool as this one because the images are not deep enough (short exposure times). 

Nevertheless, on the basis of our results, we remark that the observational strategy described in this paper can lead to the detection of metal-poor brown dwarf companions. The substellar borderline occurs at M($I$)\,=\,13.93\,mag and M($J$)\,=\,11.57\,mag for [M/H]\,=\,$-1$ and age\,=\,10\,Gyr (Baraffe et al$.$ \cite{baraffe97}; Chabrier \& Baraffe \cite{chabrier97}). For younger ages, the star--brown dwarf frontier happens at brighter magnitudes. The most massive brown dwarfs are expected to be about 1 magnitude fainter in $J$ and to show cool atmospheres with temperatures below 1000\,K; their peak of emission will take place mostly between the $R$ and $H$ wavelengths (Saumon et al$.$ \cite{saumon94}; Allard \& Hauschildt \cite{allard95}). These values are within the capabilities of mid-size telescopes and their optical and near-infrared instrumentation. As compared to solar-metallicity brown dwarfs of similar mass, models predict that low-metallicity substellar objects are significantly bluer in the infrared colors (see the turn-over of the isochrones in Figs.~\ref{jhk} and~\ref{jhkcolor}).

\begin{figure}
\centering
\includegraphics[width=8.5cm]{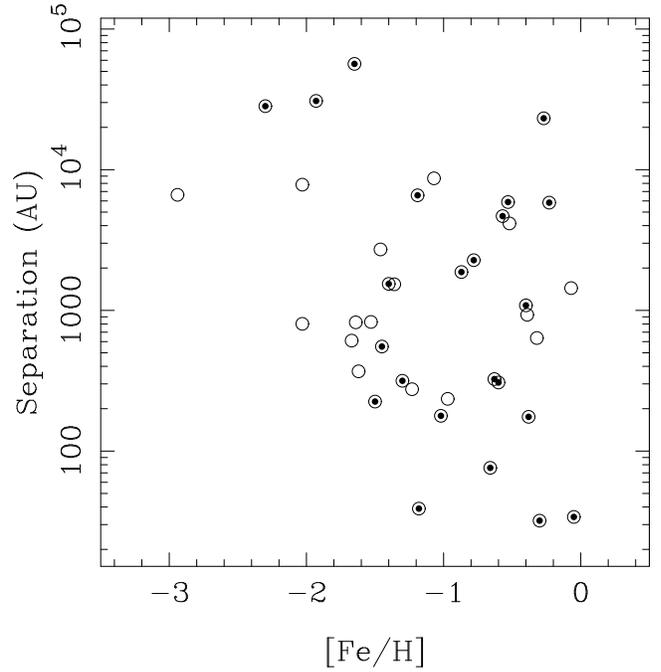}
\caption{Projected separation between companion and primary star against metallicity. Encircled dots stand for stars with Hipparcos distances.}
\label{sep} 
\end{figure}

\subsection{Wide binaries, metallicity and frequency}

Figure~\ref{sep} illustrates the projected separation of the common proper motion companions to G-, K- and M-type subdwarfs as a function of metallicity. Numerical values are provided in Tables~\ref{oldbin} and~\ref{newbin}. Separations range from a few tens of AU up to about 57000\,AU. 
The lack of small orbits at very low-metallicities is very likely an observational effect due to the distance-to-the-star distribution shown in Fig.~\ref{dist}, and should not be considered reliable. Metal-poor, low-mass stars are found in wide binary systems with orbital sizes that resemble those of solar-metallicity binary stars of similar masses (e.g., Gliese \& Jahreiss \cite{gliese88}; Poveda et al$.$ \cite{poveda94}). This result agrees with the very recent and extensive work of Latham et al$.$ (\cite{latham02}), who reported on the finding of 188 spectroscopic binaries in a sample of 1359 stars selected from the Carney-Latham proper-motion surveys. These authors concluded that there are no obvious discrepancy between the close binary characteristics in the halo and the disk populations. Concerning wide systems, we do not find significant difference in the orbital separations of metal-depleted binaries and solar-metallicity multiple stars.

About 9\%~of the metal-poor G-, K- and M-type stars in our sample have visual, common proper motion stellar companions at wide orbits ($\ge$30\,AU) and with mass ratios of 0.13--1.0. A similar binary fraction was obtained by Allen et al.~(\cite{allen00}) after examining more than 1200 low-metallicity stars from the catalog of Schuster \& Nissen (\cite{schuster89}). Other surveys of metal-deficient, high-velocity stars yielded slightly larger frequencies ($\sim$13\%), but they are all based on small statistics (e.g., Abt \& Willmarth \cite{abt87}; Mart\'\i n \& Rebolo \cite{martin92}). Poveda et al$.$ (\cite{poveda94}) provided a catalog of wide binaries and multiple stars of the solar vicinity with separations larger than 25\,AU. The great majority of the stars in their catalog (98.5\%) has solar metallicities. According to these authors, the fraction of wide binaries in the solar neighborhood is 15--20\%, which is about a factor of 2 larger than the one found for low-metallicity stars.

\begin{figure}
\centering
\includegraphics[width=8.5cm]{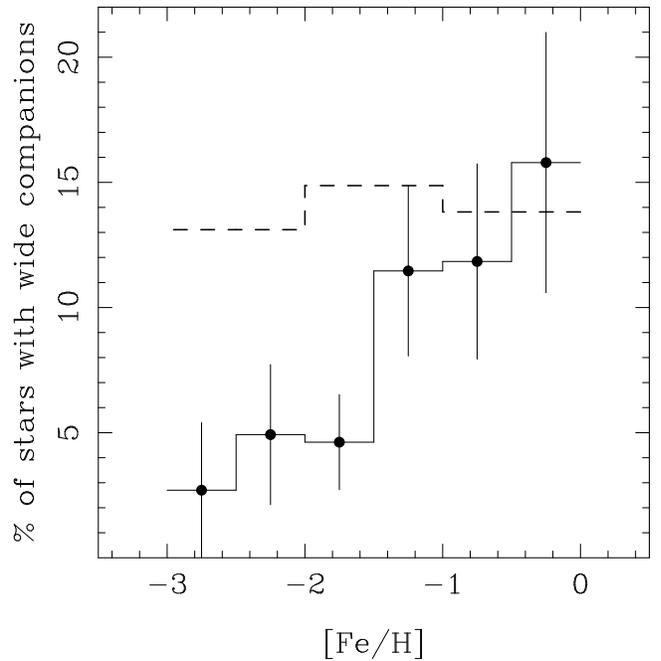}
\caption{The frequency of low-metallicity wide companions is plotted as a solid line. The bin width represents a change of 0.5\,dex in the metal logarithmic abundance. Vertical errors are Poissonian. The ``corrected'' frequency is plotted as a dashed line (see text), where the bin width is 1\,dex.}
\label{stat} 
\end{figure}

However, we will show that this ``apparent'' low frequency of metal-deficient wide companions is likely due to the undersampling of the shortest orbits at very low metallicities. The metallicity distribution of our companion stars (thick line) is compared to the distribution of the sample targets in Fig.~\ref{metal}. We calculated the cumulative metallicity distributions of both datasets and applied the Kolmogorov-Smirnov test to investigate the degree of similarity. We obtained a probability of 0.42. Such small value indicates that the two distributions are different. To further illustrate this, we evaluated the binary fraction of stars with visible companions in our survey as a function of metal content. For binary fraction, we used the number of multiple systems relative to the total number of stars per interval of metallicity. The results are depicted in Fig.~\ref{stat} (solid line). From the picture, there is a marked decrease toward very low metallicities: while 12--16\%~of the most metal-rich stars ([Fe/H]\,$\ge$\,$-1$) harbor wide companions, less than 5\%~of the most metal-deficient stars ([Fe/H]\,$\le$\,$-2$) are found in wide binaries. 

Nevertheless, Fig.~\ref{stat} needs to be corrected for the distance--metallicity relation shown in Fig.~\ref{dist}. The most metal-depleted stars in our survey are located farther away than the most metal-rich stars. Hence, we explored the widest orbits of the very low-metallicity stars, while closer regions were surveyed around more metal-rich sources. We corrected the ``direct'' binary fraction for this effect by taking into account the different orbital sizes that our survey has explored as a function of metallicity, and normalizing at [Fe/H]\,=\,[$-1$, 0]. The resulting ``corrected'' binary frequency is plotted as a dashed line in Fig.~\ref{stat}. We note that this correction is based on the assumption that companions have similar orbits throughout all metallicities. This assumption is supported by Fig.~\ref{sep}, where it is obvious that such assessment is true for the interval [Fe/H]\,=\,[$-1.5$, 0]. The corrected binary fraction is nearly flat, and suggests that 13--15\%~of the low-metallicity stars harbor wide, low-mass stellar companions with mass ratios in the range 0.13--1.0 and separations larger than 30\,AU. This overall binary frequency is in agreement with the binary fraction observed among main sequence G- to M-type stars in the solar neighbourhood (Poveda et al.~\cite{poveda94}) and young T\,Tauri stars of star-forming regions (14\,$\pm$\,1.8\%, Brandner et al.~\cite{brandner96}).

Of the total of 42 pairs, at least 11 are multiple systems ($\sim$26\%). The star G\,273-001 is resolved to be double by Hipparcos ($\rho$\,=0\farcs816, PA\,=\,226\,deg, $\Delta V$\,=\,0.22\,mag). The primary stars G\,059-032\,A (Carney et al$.$ \cite{carney94}), G\,069-004 (Carney et al$.$ \cite{carney94}), G\,061-024 (Goldberg et al$.$ \cite{goldberg02}), G\,128-064\,A (Latham et al$.$ \cite{latham02}), G\,153-067 (Carney et al$.$ \cite{carney94}), G\,173-002 (Latham et al$.$ \cite{latham02}), G\,116-009\,A (Latham et al$.$ \cite{latham02}), G\,176-046\,A (Latham et al$.$ \cite{latham92}), and G\,216-045\,A (Latham et al$.$ \cite{latham02}) are found to be spectroscopic binaries. The companion G\,095-057\,B is a single-lined system with orbital solution according to Carney et al$.$ (\cite{carney94}). The low-metallicity stars G\,009-047\,A and G\,214-001\,A also show quite large dispersions in their radial velocity measurements (Latham et al$.$ \cite{latham02}), but no conclusion on their spectroscopic binary nature can be achieved so far. Mart\'\i n et al$.$ (\cite{martin95}) noted that the new companion G\,176-046\,D is part of a multiple system formed by 4 components whose masses range from 0.13 up to 0.7\,$M_{\odot}$. The total fraction of low-metallicity multiple systems composed of three or more members cannot be derived because not all the stars have been investigated for radial velocity variations or have been imaged to very high spatial resolutions (e.g., adaptive optics). We can simply derive a lower limit from our survey, which is 2.3\%. It is very intriguing how these multiple systems formed and how they survived dynamical evolution.

We note that the 13--15\%~wide binary frequency observed among low-metallicity G- to M-type stars is also comparable to the fraction of close (unresolved) binaries, which is $\sim$14\%~according to the spectroscopic work by Latham et al$.$ (\cite{latham02}). Furthermore, many of the low-mass companions are separated from their primary stars by more than 100\,AU. If metal-depleted double or multiple systems are formed at birth, this suggests that they were not summoned to the very severe dense conditions of globular clusters for very long (Ryan \cite{ryan92}). On the contrary, their existence in significant numbers suggests that they were likely formed in more relaxed environments. The similarity between the incidence of binaries among low-metallicities and solar-metallicities also suggests that the chemical composition of the parental molecular cloud has little influence over the stellar formation of wide binary and multiple systems. This may also apply to substellar formation. Solar-metallicity brown dwarfs are found as companions to stars in a great variety of orbital sizes: from a few tens of AU (e.g., Nakajima et al$.$ \cite{nakajima95}), to several hundred AU (e.g., Rebolo et al$.$ \cite{rebolo98}), and to several thousand AU (e.g., Burgasser et al$.$ \cite{burgasser00}). The search for metal-poor substellar counterparts is the natural extension of the present work.

\section{Summary and conclusions}

We explored the nearby regions around 473 low-metallicity G- to M-type stars searching for low-mass stellar companions that are orbiting their primary stars at wide separations (typically $\ge$30\,AU). The great majority of the target stars, with high proper motions and metallicities in the range [Fe/H]\,=\,[$-$3.5, 0.0], were selected from Carney et al$.$ (\cite{carney94}) and Laird et al$.$ (\cite{laird88}). All of them were imaged in the $I$-band with a 1-m class telescope. Optical $V$ and $R$ data were also collected for a large number of the targets. The dynamical range of the {\sc ccd} detectors allowed us to detect companions up to 5 magnitudes fainter (completeness) than the target star. The physical link between a candidate and its primary star was assessed by means of photometric and proper motion measurements. We also searched the literature and various archives to complete our survey. We identified 13 new proper motion companions and retrieved 29 previously known companions. This suggests that about one third of the low-mass companions are missing in previous proper motion searches. 

2MASS $JHK_s$ photometry is provided for a total of 39 companions out of 42. On the basis of optical and near-infrared colors, the 13 new companions are dwarfs. Two out of the 29 previously known companions are white dwarfs. We produced optical and near-infrared color-magnitude and color-color diagrams, and overplotted low-metallicity models from Baraffe et al$.$ (\cite{baraffe97}). The agreement between observations and theory is reasonable, indicating that dwarf companions have masses in the range 0.1--0.5\,$M_{\odot}$. 

Low-resolution optical spectra from 386 to 1000\,nm were obtained for 8 of the new proper motion companions, for which we derived subdwarf spectral types esdK7.2--sdM3.0 (error bar of half a subclass). The spectra of significantly metal-poor companions are dominated by strong MgH and CaH molecular absorptions. At the resolution of our data, spectra appear featureless redward of 800\,nm, except for the CaII triplet that remains detectable down to esdM3. The molecular features between 600 and 800\,nm are quite sensitive to temperature changes in contrast to the blue spectra.

Proper motion pairs have projected separations between $\sim$30 and $\sim$57000\,AU. Very wide companions are also identified among the most metal-depleted stars. These orbital sizes are similar to those of solar-metallicity binaries and multiple stars. After correcting for the effect of increasing distance to the most metal-deficient stars in our survey, we determined that 13--15\%~of the low-metallicity G- to M-type stars harbor wide companions. This binary frequency is very similar to the binary fraction observed among stars of the solar vicinity and T\,Tauri stars of star-forming regions, suggesting that metallicity is not a key parameter in the stellar formation of wide double and multiple systems.

\begin{acknowledgements}
We thank R$.$ Rebolo, M$.$ Guerrero, E$.$ Oblak, and J$.$ Webb for
their help with the observations. This publication makes use of data
products from the Two Micron All Sky Survey (2MASS), which is a joint
project of the University of Massachusetts and the Infrared Processing
and Analysis Center/California Institute of Technology, funded by the
National Aeronautics and Space Administration and the National Science
Foundation. This research has made use of the SIMBAD database,
operated at CDS, Strasbourg, France. It has also made use of the
Digitized Sky Surveys produced at the Space Telescope Science
Institute under U.S$.$ Government grant NAG W-2166. This work is
partly financed by the Spanish projects AYA2003-053555 and Ram\'on y
Cajal.
\end{acknowledgements}

\include{table01}

\include{table02}

\include{table03}

\include{table04}

\include{table05}
\include{table06}

\include{table07}

\include{table08}

\include{table09}

\include{table10}

\include{table11}

\include{table12}

\include{table13}

\include{table14}

\include{table15}

\include{table16}

\include{table17}

\include{table18}

\include{table19}

\include{table20}

\include{table21}

\include{table22}

\include{table23}

\include{table24}
\include{table25}

\end{document}

%% file: table01.tex
\setcounter{table}{0}
\begin{table*}
\caption{List of photometric and astrometric observations.\label{photlog}}
\begin{center}
\begin{tabular}{lrccrccrrrc}
\hline
\hline
Object      &$V_p$~ & Fil. & Tel.  & Exp. & FWHM    & Date & $\rho$~ & PA~ & $\Delta$mag & Notes \\
            &       &      &       & (s)  &(\arcsec)&      &(\arcsec)&(deg)&             &       \\
\hline		
G\,001-004    & 10.69 & $I$ & IAC80      &   35  &  1.9     & 01/08/94  & $\ge$25 &         &        &  \\
G\,001-012    & 13.31 & $I$ & IAC80      &  100  &  2.6     & 02/08/94  & 23.29   &   276.8 & 4.49   & 1\\
G\,001-013    & 15.23 & $I$ & IAC80      &  330  &  2.5     & 02/08/94  & 20.48   &   341.2 & 3.89   & 1\\
G\,001-028    & 10.50 & $I$ & IAC80      &   15  &  1.4     & 03/08/94  & $\ge$20 &         &        &  \\
G\,004-037    & 11.43 & $I$ & IAC80      &  120  &  2.3     & 15/02/94  & $\ge$25 &         &        &  \\
              &       & $I$ &            &  180  &  2.2     &           & $\ge$25 &         &        &  \\
\end{tabular}
\end{center}
\end{table*}

%% file: table02.tex
\setcounter{table}{0}
\begin{table*}
\caption{{\sl Continued ...}}
\begin{center}
\begin{tabular}{lrccrccrrrc}
\hline
\hline
Object      &$V_p$~ & Fil. & Tel.  & Exp. & FWHM    & Date & $\rho$~ & PA~ & $\Delta$mag & Notes \\
            &       &      &       & (s)  &(\arcsec)&      &(\arcsec)&(deg)&             &        \\
\hline		
G\,016-020    & 10.80 & $I$ & IAC80      &  40   & 1.5      & 07/06/95  & $\ge$25 &         &        &  \\
G\,016-025    & 13.34 & $I$ & IAC80      & 500   & 1.6      & 07/06/95  & $\ge$25 &         &        &  \\
G\,016-035    & 13.12 & $I$ & IAC80      & 180   & 2.1      & 16/02/94  & $\ge$25 &         &        &  \\
              &       & $V$ &            & 240   & 2.3      &           & $\ge$25 &         &        &  \\
G\,017-025$*$ &  9.63 & $I$ & IAC80      &  10   & 1.5      & 06/06/95  & $\ge$25 &         &        &  \\
G\,017-037    & 13.55 & $I$ &   JKT      &  60   & 3.2      & 18/03/91  &  5.24   &  215.5  &  2.69  & 2,3\\
\end{tabular}
\end{center}
\end{table*}

%% file: table03.tex
\setcounter{table}{0}
\begin{table*}
\caption{{\sl Continued ...}}
\begin{center}
\begin{tabular}{lrccrccrrrc}
\hline
\hline
Object      &$V_p$~ & Fil. & Tel.  & Exp. & FWHM    & Date & $\rho$~ & PA~ & $\Delta$mag & Notes \\
            &       &      &       & (s)  &(\arcsec)&      &(\arcsec)&(deg)&             &       \\
\hline		
              &       &     &            &       &          &           & 20.45   &   67.8  &  6.59  & 1,2\\
              &       &     &            &       &          &           & 22.26   &  195.8  &  7.10  & 1\\
              &       & $V$ &            & 100   & 2.0      & 27/07/94  &   ---   &    ---  &   ---  &  \\
              &       &     &            &       &          &           & 13.71   &  290.0  &  6.49  & 1,2\\
              &       &     &            &       &          &           & 18.66   &  174.6  &  5.99  & 1,2\\
              &       &     &            &       &          &           & 20.4:   &   68:   &  6.8:  & 1,2\\
\end{tabular}
\end{center}
\end{table*}

%% file: table04.tex
\setcounter{table}{0}
\begin{table*}
\caption{{\sl Continued ...}}
\begin{center}
\begin{tabular}{lrccrccrrrc}
\hline
\hline
Object      &$V_p$~ & Fil. & Tel.  & Exp. & FWHM    & Date & $\rho$~ & PA~ & $\Delta$mag & Notes \\
            &       &      &       & (s)  &(\arcsec)&      &(\arcsec)&(deg)&             &       \\
\hline		
              &       & $R$ &            &  12   & 2.0      &           & 23.50   &  308.7  &  6.92  & 2\\
              &       &     &            &       &          &           &   ---   &    ---  &    --- & 1\\
G\,025-031    & 10.65 & $I$ &   IAC80    &  80   & 2.0      & 30/07/94  & $\ge$25 &         &        &  \\
              &       & $R$ &            &  80   & 2.0      &           & $\ge$25 &         &        &  \\
G\,026-001    & 11.27 & $I$ &   IAC80    & 150   & 1.3      & 07/06/95  & $\ge$25 &         &        &  \\
G\,026-008    & 10.47 & $I$ &   IAC80    &  10   & 1.8      & 03/08/94  &  8.54   &  306.3  &  3.26  & 4\\
\end{tabular}
\end{center}
\end{table*}

%% file: table05.tex
\setcounter{table}{0}
\begin{table*}
\caption{{\sl Continued ...}}
\begin{center}
\begin{tabular}{lrccrccrrrc}
\hline
\hline
Object      &$V_p$~ & Fil. & Tel.  & Exp. & FWHM    & Date & $\rho$~ & PA~ & $\Delta$mag & Notes \\
            &       &      &       & (s)  &(\arcsec)&      &(\arcsec)&(deg)&             &       \\
\hline		
              &       & $I$ &            &  60   & 2.3      & 21/08/96  &  7.00   &  177.8  &  5.18  & 3\\
              &       & $R$ &            &  60   & 2.2      &           &  6.97   &  179.1  &  5.52  & 3\\
              &       & $I$ &            & 120   & 2.1      & 22/08/96  &  7.00   &  179.1  &  5.11  & 3\\
              &       & $V$ &            & 120   & 2.1      &           &  6.96   &  178.8  &  6.10  & 3\\
G\,029-075    & 13.42 & $I$ &   IAC80    & 230   & 2.1      & 02/08/94  & $\ge$25 &         &        &  \\
G\,030-024    & 10.56 & $I$ &   IAC80    &  45   & 2.2      & 30/07/94  & $\ge$25 &         &        &  \\
\end{tabular}
\end{center}
\end{table*}

%% file: table06.tex
\setcounter{table}{0}
\begin{table*}
\caption{{\sl Continued ...}}
\begin{center}
\begin{tabular}{lrccrccrrrc}
\hline
\hline
Object      &$V_p$~ & Fil. & Tel.  & Exp. & FWHM    & Date & $\rho$~ & PA~ & $\Delta$mag & Notes \\
            &       &      &       & (s)  &(\arcsec)&      &(\arcsec)&(deg)&             &       \\
\hline		
G\,059-027    & 10.90 & $I$ &   IAC80    & 180   & 1.8      & 07/01/94  & $\ge$25 &         &        &  \\
G\,059-032    & 9.04  & $I$ &   IAC80    &  60   & 2.0      & 26/01/94  & 13.70   &   71.1  &  8.3:  & 2\\
              &       & $R$ &            &  30   & 1.7      &           & 13.24   &   71.0  &  8.4:  & 2\\
              &       & $V$ &            &  60   & 2.0      &           &  ---    &    ---  &   ---  & 2\\
G\,060-026    & 9.81  & $I$ &   JKT      &  20   & 2.6      & 19/03/91  & $\ge$25 &         &        &  \\
              &       & $R$ &   JKT      &  20   & 2.7      &           & $\ge$25 &         &        &  \\
\end{tabular}
\end{center}
\end{table*}

%% file: table07.tex
\setcounter{table}{0}
\begin{table*}
\caption{{\sl Continued ...}}
\begin{center}
\begin{tabular}{lrccrccrrrc}
\hline
\hline
Object      &$V_p$~ & Fil. & Tel.  & Exp. & FWHM    & Date & $\rho$~ & PA~ & $\Delta$mag & Notes \\
            &       &      &       & (s)  &(\arcsec)&      &(\arcsec)&(deg)&             &       \\
\hline		
G\,067-038    & 8.38  & $I$ &   IAC80    &  10   & 1.7      & 01/08/94  & 22.7::  &   91::  &  8.4:: & 1\\
G\,068-003    & 9.74  & $I$ &   IAC80    &  30   & 2.1      & 30/07/94  & $\ge$25 &         &        &  \\
G\,068-010    & 8.96  & $I$ &   IAC80    &  15   & 1.7      & 01/08/94  & $\ge$25 &         &        &  \\
G\,068-029    & 10.88 & $I$ &   IAC80    &  85   & 2.5      & 30/07/94  & $\ge$25 &         &        &  \\
G\,068-030    & 9.08  & $I$ &   IAC80    &  30   & 2.5      & 30/07/94  & $\ge$25 &         &        &  \\
G\,068-031    & 7.66  & $I$ &   IAC80    &   5   & 2.0      & 02/08/94  & $\ge$25 &         &        &  \\
\end{tabular}
\end{center}
\end{table*}

%% file: table08.tex
\setcounter{table}{0}
\begin{table*}
\caption{{\sl Continued ...}}
\begin{center}
\begin{tabular}{lrccrccrrrc}
\hline
\hline
Object      &$V_p$~ & Fil. & Tel.  & Exp. & FWHM    & Date & $\rho$~ & PA~ & $\Delta$mag & Notes \\
            &       &      &       & (s)  &(\arcsec)&      &(\arcsec)&(deg)&             &       \\
\hline		
              &       &     &            &       &          &           & 16.33   &  320.8  &  6.83  &  \\
              &       &     &            &       &          &           & 19.76   &   69.1  &  7.31  &  \\
              &       &     &            &       &          &           & 19.25   &  236.6  &  5.42  & 2,3\\
              &       & $I$ &            & 120   & 3.2      & 18/11/02  &  8.17   &   55.6  &  5.02  & 2,3\\
              &       &     &            &       &          &           &  9.29   &   18.6  &  6.57  &  \\
              &       &     &            &       &          &           &  8.20   &   91.0  &  7.07  &  \\
\end{tabular}
\end{center}
\end{table*}

%% file: table09.tex
\setcounter{table}{0}
\begin{table*}
\caption{{\sl Continued ...}}
\begin{center}
\begin{tabular}{lrccrccrrrc}
\hline
\hline
Object      &$V_p$~ & Fil. & Tel.  & Exp. & FWHM    & Date & $\rho$~ & PA~ & $\Delta$mag & Notes \\
            &       &      &       & (s)  &(\arcsec)&      &(\arcsec)&(deg)&             &       \\
\hline		
              &       & $I$ &            & 180   & 1.6      &           & $\ge$25 &         &        &  \\
G\,115-049    & 11.60 & $I$ &   JKT      &  30   & 1.2      & 17/01/91  & 23.97   &  81.5   &  3.30  & 1\\
              &       & $V$ &            &  30   & 1.0      &           & 23.93   &  81.4   &  4.61  & 1\\
G\,115-058    & 12.08 & $I$ &   IAC80    & 300   & 2.2      & 17/02/94  & $\ge$25 &         &        &  \\
              &       & $V$ &            & 300   & 2.2      &           & $\ge$25 &         &        &  \\
G\,115-062    & 13.74 & $I$ &   JKT      &  30   & 2.2      & 20/03/91  & $\ge$25 &         &        &  \\
\end{tabular}
\end{center}
\end{table*}

%% file: table10.tex
\setcounter{table}{0}
\begin{table*}
\caption{{\sl Continued ...}}
\begin{center}
\begin{tabular}{lrccrccrrrc}
\hline
\hline
Object      &$V_p$~ & Fil. & Tel.  & Exp. & FWHM    & Date & $\rho$~ & PA~ & $\Delta$mag & Notes \\
            &       &      &       & (s)  &(\arcsec)&      &(\arcsec)&(deg)&             &       \\
\hline		
G\,124-030    & 15.45 & $I$ &   IAC80    & 800   & 2.1      & 06/06/95  & 20.91   &  163.6  &  0.11  &  \\
G\,124-035    & 14.81 & $I$ &   IAC80    & 800   & 1.8      & 07/06/95  & 12.71   &  244.6  &  3.52  & 1\\
G\,125-039    & 14.52 & $I$ &   IAC80    & 600   & 1.8      & 07/06/95  &  4.25   &  139.2  &  2.64  & 1\\
              &       &     &            &       &          &           &  6.52   &   35.4  &  3.98  & 1\\
              &       &     &            &       &          &           &  9.18   &  254.5  &  4.37  &  \\
              &       &     &            &       &          &           &  9.88   &  106.2  &  3.89  &  \\
\end{tabular}
\end{center}
\end{table*}

%% file: table11.tex
\setcounter{table}{0}
\begin{table*}
\caption{{\sl Continued ...}}
\begin{center}
\begin{tabular}{lrccrccrrrc}
\hline
\hline
Object      &$V_p$~ & Fil. & Tel.  & Exp. & FWHM    & Date & $\rho$~ & PA~ & $\Delta$mag & Notes \\
            &       &      &       & (s)  &(\arcsec)&      &(\arcsec)&(deg)&             &       \\
\hline		
              &       &     &            &       &          &           & 18.55   &  306.4  &  5.10  & 2,3\\
G\,128-080    & 11.99 & $I$ &   IAC80    &  80   & 2.4      & 29/07/94  & 13.96   &   11.4  &  4.74  & 1,2\\
              &       & $R$ &            &  80   & 2.3      &           & 13.94   &   11.5  &  4.72  & 1,2\\
G\,128-083    & 7.91  & $I$ &   IAC80    &   5   & 2.4      & 02/08/94  & $\ge$25 &         &        &  \\
G\,128-084    & 10.12 & $I$ &   IAC80    &  45   & 2.3      & 30/07/94  & $\ge$25 &         &        &  \\
G\,129-042    & 8.94  & $I$ &   IAC80    &  20   & 2.1      & 30/07/94  & $\ge$25 &         &        &  \\
\end{tabular}
\end{center}
\end{table*}

%% file: table12.tex
\setcounter{table}{0}
\begin{table*}
\caption{{\sl Continued ...}}
\begin{center}
\begin{tabular}{lrccrccrrrc}
\hline
\hline
Object      &$V_p$~ & Fil. & Tel.  & Exp. & FWHM    & Date & $\rho$~ & PA~ & $\Delta$mag & Notes \\
            &       &      &       & (s)  &(\arcsec)&      &(\arcsec)&(deg)&             &       \\
\hline		
              &       & $I$ &   IAC80    & 100   & 2.0      & 27/08/94  & 21.55   &  134.9  &  7.79  & 1\\
G\,142-044    & 11.15 & $I$ &   IAC80    &  20   & 1.9      & 15/07/95  &  7.0:   &   10:   &  6:    &  \\
              &       &     &            &       &          &           &  9.93   &   97.6  &  5.70  & 1\\
              &       &     &            &       &          &           & 14.12   &  213.3  &  5.05  & 1\\
              &       &     &            &       &          &           & 15.1:   &  162:   &  5.9:  & 1\\
              &       &     &            &       &          &           & 17.95   &  335.8  &  6.17  & 1\\
\end{tabular}
\end{center}
\end{table*}

%% file: table13.tex
\setcounter{table}{0}
\begin{table*}
\caption{{\sl Continued ...}}
\begin{center}
\begin{tabular}{lrccrccrrrc}
\hline
\hline
Object      &$V_p$~ & Fil. & Tel.  & Exp. & FWHM    & Date & $\rho$~ & PA~ & $\Delta$mag & Notes \\
            &       &      &       & (s)  &(\arcsec)&      &(\arcsec)&(deg)&             &       \\
\hline		
              &       &     &            &       &          &           & 22.86   &  207.7  &  7.52  &  \\
G\,156-007    & 11.82 & $I$ &   IAC80    &  20   & 2.0      & 15/07/95  & $\ge$25 &         &        &  \\
G\,157-021    & 9.27  & $I$ &   IAC80    &  10   & 1.3      & 03/08/94  & $\ge$25 &         &        &  \\
G\,157-072    & 15.72 & $I$ &   IAC80    &  30   & 1.4      & 03/08/95  & $\ge$25 &         &        &  \\
G\,157-093    & 10.12 & $I$ &   IAC80    &  15   & 1.4      & 03/08/94  & $\ge$25 &         &        &  \\
G\,158-021    & 8.59  & $I$ &   IAC80    &   5   & 1.1      & 03/08/94  & $\ge$25 &         &        &  \\
\end{tabular}
\end{center}
\end{table*}

%% file: table14.tex
\setcounter{table}{0}
\begin{table*}
\caption{{\sl Continued ...}}
\begin{center}
\begin{tabular}{lrccrccrrrc}
\hline
\hline
Object      &$V_p$~ & Fil. & Tel.  & Exp. & FWHM    & Date & $\rho$~ & PA~ & $\Delta$mag & Notes \\
            &       &      &       & (s)  &(\arcsec)&      &(\arcsec)&(deg)&             &       \\
\hline		
              &       & $R$ &            & 100   & 1.8      &           &  7.23   &   27.9  &  0.78  & 1\\
G\,171-050    & 13.08 & $I$ &   IAC80    & 200   & 2.9      & 02/08/94  &  8.6:   &    7:   &  5.0:  & 2,3\\
              &       &     &            &       &          &           & 22.44   &  268.3  &  4.06  & 4\\
              &       & $I$ &            & 290   & 2.6      &           &  8.69   &    7.5  &  4.30  & 2,3\\
              &       &     &            &       &          &           & 22.48   &  268.8  &  4.26  & 4\\
              &       & $I$ &            &  40   &          & 17/07/95  &  8.5:   &    6:   &  5.1:  & 2,3\\
\end{tabular}
\end{center}
\end{table*}

%% file: table15.tex
\setcounter{table}{0}
\begin{table*}
\caption{{\sl Continued ...}}
\begin{center}
\begin{tabular}{lrccrccrrrc}
\hline
\hline
Object      &$V_p$~ & Fil. & Tel.  & Exp. & FWHM    & Date & $\rho$~ & PA~ & $\Delta$mag & Notes \\
            &       &      &       & (s)  &(\arcsec)&      &(\arcsec)&(deg)&             &       \\
\hline		
G\,178-056    & 11.90 & $R$ &   JKT      &  20   & 1.2      & 20/03/91  &  7.80   &  181.8  &  2.13  & 2,3\\
              &       & $I$ &            &  60   & 1.6      & 05/05/93  &  7.38   &  177.6  &  2.13  & 2,3\\
              &       & $V$ &            & 100   & 1.6      &           &  7.39   &  177.6  &  2.12  & 2,3\\
              &       & $I$ &   NOT      &  10   & 1.5      & 02/05/94  &  7.18   &  179.2  &  2.16  & 2,3\\
              &       & $I$ &            &  25   & 0.8      &           &  7.18   &  179.1  &  2.14  & 2,3\\
              &       & $V$ &            &  30   & 0.8      &           &  7.18   &  179.1  &  2.15  & 2,3\\
\end{tabular}
\end{center}
\end{table*}

%% file: table16.tex
\setcounter{table}{0}
\begin{table*}
\caption{{\sl Continued ...}}
\begin{center}
\begin{tabular}{lrccrccrrrc}
\hline
\hline
Object      &$V_p$~ & Fil. & Tel.  & Exp. & FWHM    & Date & $\rho$~ & PA~ & $\Delta$mag & Notes \\
            &       &      &       & (s)  &(\arcsec)&      &(\arcsec)&(deg)&             &       \\
\hline		
              &       &     &            &       &          &           & 23.19   &  140.5  &  6.03  & 3\\
              &       & $R$ &            & 200   & 2.0      & 28/07/94  &  6.5:   &  131:   &  6.1:  & 3\\
              &       &     &            &       &          &           & 10.86   &   30.4  &  5.02  & 3\\
              &       &     &            &       &          &           & 16.55   &  245.0  &  5.51  & 3\\
              &       &     &            &       &          &           & 21.67   &   49.0  &  4.47  & 3\\
              &       &     &            &       &          &           & 23.26   &  140.2  &  5.94  & 3\\
\end{tabular}
\end{center}
\end{table*}

%% file: table17.tex
\setcounter{table}{0}
\begin{table*}
\caption{{\sl Continued ...}}
\begin{center}
\begin{tabular}{lrccrccrrrc}
\hline
\hline
Object      &$V_p$~ & Fil. & Tel.  & Exp. & FWHM    & Date & $\rho$~ & PA~ & $\Delta$mag & Notes \\
            &       &      &       & (s)  &(\arcsec)&      &(\arcsec)&(deg)&             &       \\
\hline		
              &       &     &            &       &          &           & 19.6:   &  108:   &  5.9:  & 1\\
G\,188-022    & 10.05 & $I$ &   IAC80    &  30   & 2.4      & 31/07/94  &  5.02   &  306.9  &  5.54  & 5\\
              &       & $I$ &            &  50   & 2.5      &           &  5.12   &  309.6  &  5.45  & 5\\
              &       & $I$ &            &  50   & 2.6      &           &  5.09   &  308.4  &  5.43  & 5\\
              &       & $I$ &            &  30   & 2.6      & 30/10/02  &  5.72   &  305.2  &  5.39  & 5\\
              &       & $R$ &            &  15   & 2.6      &           &  5.58   &  305.7  &  6.21  & 5\\
\end{tabular}
\end{center}
\end{table*}

%% file: table18.tex
\setcounter{table}{0}
\begin{table*}
\caption{{\sl Continued ...}}
\begin{center}
\begin{tabular}{lrccrccrrrc}
\hline
\hline
Object      &$V_p$~ & Fil. & Tel.  & Exp. & FWHM    & Date & $\rho$~ & PA~ & $\Delta$mag & Notes \\
            &       &      &       & (s)  &(\arcsec)&      &(\arcsec)&(deg)&             &       \\
\hline		
              &       & $I$ &            &  40   & 2.3      & 06/05/94  & 17.88   &   64.7  &  7.7:  & 1\\
G\,206-001    & 12.44 & $I$ &   IAC80    & 150   & 1.4      & 06/06/95  & 20.36   &   16.0  &  7.71  & 1\\
              &       &     &            &       &          &           & 23.94   &  109.8  &  5.37  & 1\\
G\,206-020    & 12.86 & $I$ &   IAC80    & 150   & 1.6      & 06/06/95  & 16.74   &  115.7  &  6.08  & 1\\
              &       &     &            &       &          &           & 17.84   &    4.7  &  6.12  & 1\\
              &       &     &            &       &          &           & 23.54   &  264.0  &  6.24  & 1\\
\end{tabular}
\end{center}
\end{table*}

%% file: table19.tex
\setcounter{table}{0}
\begin{table*}
\caption{{\sl Continued ...}}
\begin{center}
\begin{tabular}{lrccrccrrrc}
\hline
\hline
Object      &$V_p$~ & Fil. & Tel.  & Exp. & FWHM    & Date & $\rho$~ & PA~ & $\Delta$mag & Notes \\
            &       &      &       & (s)  &(\arcsec)&      &(\arcsec)&(deg)&             &       \\
\hline		
G\,212-055    & 14.15 & $I$ &   IAC80    & 100   & 2.1      & 29/07/94  &  6.80   &  328.3  &  3.94  & 3\\
              &       &     &            &       &          &           & 15.60   &   60.8  &  4.16  & 1\\
              &       &     &            &       &          &           & 22.7:   &   18:   &  4.5:  & 1\\
              &       & $I$ &            & 100   & 2.1      & 17/07/95  &  7.05   &  328.6  &  4.04  & 3\\
G\,214-001    & 12.08 & $I$ &   IAC80    & 100   & 2.0      & 29/07/94  &  5.15   &  352.8  &  4.59  & 5\\
              &       &     &            &       &          &           & 15.63   &   77.1  &  5.47  & 2,3\\
\end{tabular}
\end{center}
\end{table*}

%% file: table20.tex
\setcounter{table}{0}
\begin{table*}
\caption{{\sl Continued ...}}
\begin{center}
\begin{tabular}{lrccrccrrrc}
\hline
\hline
Object      &$V_p$~ & Fil. & Tel.  & Exp. & FWHM    & Date & $\rho$~ & PA~ & $\Delta$mag & Notes \\
            &       &      &       & (s)  &(\arcsec)&      &(\arcsec)&(deg)&             &       \\
\hline		
G\,216-045    & 11.24 & $I$ &   IAC80    &  70   & 1.9      & 29/07/94  &  3.7:   &   97:   &  2.9:  & 3\\
              &       &     &            &       &          &           & 12.54   &  190.6  &  6.71  & 3\\
              &       &     &            &       &          &           & 14.8:   &   42:   &  6.9:  & 3\\
              &       &     &            &       &          &           & 19.45   &  111.3  &  6.95  & 3\\
              &       &     &            &       &          &           & 31.21   &   44.1  &  6.18  & 5\\
              &       & $I$ &            & 100   & 2.0      & 16/10/03  &   ---   &    ---  &   ---  & 3\\
\end{tabular}
\end{center}
\end{table*}


%% file: table21.tex
\setcounter{table}{0}
\begin{table*}
\caption{{\sl Continued ...}}
\begin{center}
\begin{tabular}{lrccrccrrrc}
\hline
\hline
Object      &$V_p$~ & Fil. & Tel.  & Exp. & FWHM    & Date & $\rho$~ & PA~ & $\Delta$mag & Notes \\
            &       &      &       & (s)  &(\arcsec)&      &(\arcsec)&(deg)&             &       \\
\hline		
              &       &     &            &       &          &           & 25.00   &  228.9  &  3.34  & 3\\
              &       & $R$ &            &  40   & 2.1      & 01/08/94  & 11.64   &  267.6  &  4.49  & 4\\
              &       &     &            &       &          &           & 24.99   &  228.9  &  3.77  & 3\\
              &       & $I$ &            &  20   & 1.8      & 17/07/95  & 11.62   &  266.7  &  3.63  & 4\\
              &       &     &            &       &          &           & 24.87   &  228.8  &  3.38  & 3\\
              &       & $-$ &            &  50   & 1.9      & 22/08/96  & 11.61   &  267.3  &   ---  & 4\\
\end{tabular}
\end{center}
\end{table*}

%% file: table22.tex
\setcounter{table}{0}
\begin{table*}
\caption{{\sl Continued ...}}
\begin{center}
\begin{tabular}{lrccrccrrrc}
\hline
\hline
Object      &$V_p$~ & Fil. & Tel.  & Exp. & FWHM    & Date & $\rho$~ & PA~ & $\Delta$mag & Notes \\
            &       &      &       & (s)  &(\arcsec)&      &(\arcsec)&(deg)&             &       \\
\hline		
              &       & $I$ &   IAC80    &  60   & 1.7      & 06/05/94  & 23.59   &  174.7  &  5.57  & 1\\
G\,241-004    & 12.98 & $I$ &   IAC80    & 100   & 3.0      & 30/07/94  & 14.69   &  189.6  &  2.94  & 1\\
              &       &     &            &       &          &           & 17.2:   &  332:   &  5.2:  & 1\\
              &       &     &            &       &          &           & 20.30   &  144.4  &  3.76  & 1\\
              &       & $R$ &            & 100   & 3.2      & 30/07/94  & 14.69   &  189.7  &  3.17  & 1\\
              &       &     &            &       &          &           & 17.2:   &  332:   &  5.2:  & 1\\
\end{tabular}
\end{center}
\end{table*}

%% file: table23.tex
\setcounter{table}{0}
\begin{table*}
\caption{{\sl Continued ...}}
\begin{center}
\begin{tabular}{lrccrccrrrc}
\hline
\hline
Object      &$V_p$~ & Fil. & Tel.  & Exp. & FWHM    & Date & $\rho$~ & PA~ & $\Delta$mag & Notes \\
            &       &      &       & (s)  &(\arcsec)&      &(\arcsec)&(deg)&             &       \\
\hline		
G\,261-029    & 7.81  & $I$ &   IAC80    &  10   & 2.0      & 29/07/94  & $\ge$25 &         &        &  \\
G\,261-036    & 8.65  & $I$ &   IAC80    &  20   & 2.0      & 29/07/94  & $\ge$25 &         &        &  \\
              &       & $R$ &            &  20   & 2.0      &           & $\ge$25 &         &        &  \\
G\,261-038    & 8.70  & $I$ &   IAC80    &  10   & 2.5      & 29/07/94  & 17.6:   &  172:   &  7.5:  & 1\\
              &       & $R$ &            &  10   & 2.0      &           & 17.5:   &  172:   &  8:    & 1\\
G\,262-014    & 11.46 & $I$ &   IAC80    & 260   & 3.3      & 29/05/94  & $\ge$25 &         &        &  \\
\end{tabular}
\end{center}
\end{table*}

%% file: table24.tex
\setcounter{table}{0}
\begin{table*}
\caption{{\sl Continued ...}}
\begin{center}
\begin{tabular}{lrccrccrrrc}
\hline
\hline
Object      &$V_p$~ & Fil. & Tel.  & Exp. & FWHM    & Date & $\rho$~ & PA~ & $\Delta$mag & Notes \\
            &       &      &       & (s)  &(\arcsec)&      &(\arcsec)&(deg)&             &       \\
\hline		
LHS\,2500     & 18.90 & $I$ &  JKT       & 600   & 1.4      & 17/01/91  & 6.3:    &   90:   &  3.2:  & 2,3\\
              &       & $I$ &   NOT      & 240   & 1.0      & 02/05/94  & 8.96    &   91.8  &  3.52  & 2,3\\
              &       & $V$ &            & 350   & 1.2      &           & 8.9:    &   91:   &  3.64  & 2,3\\
LHS\,3780     & 11.85 & $I$ &   IAC80    & 200   & 1.4      & 07/06/95  & 7.13    &  154.2  &  5.66  & 1\\ 
LTT\,5454     & 12.42 & $I$ &   JKT      & 60    & 2.5      & 19/03/91  & $\ge$25 &         &        &  \\
              &       & $R$ &            & 60    & 2.7      &           & $\ge$25 &         &        &  \\
\end{tabular}
\end{center}
{\sc notes to the last column.--}\\
1~No common proper motion companion after comparison with POSS2 red plates.\\
2~No ``color'' companion.\\
3~No proper motion companion.\\
4~Previously known common proper motion companion (see Table~\ref{oldbin}).\\
5~Newly identified common proper motion companion (see Table~\ref{newbin}).\\
\end{table*}

%% file: table25.tex
\begin{table*}
\smallskip
{\sc notes on individual objects.--}\\
\begin{tabular}{ll}
G\,014-005: & One companion (CCDM\,J12497+0112\,B, $V$\,=\,9.6) at 1\farcm85 and PA\,=\,340\,deg. \\
            & We do not confirm it using our data and images from the Digitized Sky Surveys.\\
G\,017-025: & One companion (CCDM\,J16348-0413\,C, $V$\,=\,13.5) at 1\farcm55 and PA\,=\,321\,deg.\\
            & We do not confirm it using our data and images from the Digitized Sky Surveys.\\
G\,026-009: & Known proper motion companion (G\,026-010, $V$\,=\,14.58) at 2\farcm2 and PA\,=\,29\,deg. According to \\
            & Allen et al$.$ (\cite{allen00}), there is a brighter companion at 0\farcs7.\\
G\,028-043: & One companion (CCDM\,J23096+0043\,B, $V$\,=\,11.5) at 12\farcs2 and PA\,=\,195\,deg.\\
            & We do not confirm it using our data and images from the Digitized Sky Surveys.\\
\end{tabular}
\end{table*}